\documentclass[12 pt]{article}
\usepackage[utf8]{inputenc}
\usepackage[top=1.5cm, bottom=2cm, left=2.5cm, right=2cm]{geometry}
\usepackage{amsmath}
\usepackage{physics}
\usepackage{graphicx}
\usepackage{subcaption}
\usepackage{cite}

\newcommand{\be}{\begin{equation}}
\newcommand{\ee}{\end{equation}}
\newcommand{\ba}{\begin{align}}
\newcommand{\ea}{\end{align}}
\newcommand{\bg}{\begin{gather}}
\newcommand{\eg}{\end{gather}}
\newcommand{\bseq}{\begin{subequations}}
\newcommand{\eseq}{\end{subequations}}

\renewcommand{\tanh}{\mathop{\rm th}\nolimits}

\renewcommand{\ln}{\mathop{\rm ln}\nolimits}

\renewcommand{\Im}{\mathop{\rm Im}\nolimits}
\renewcommand{\Re}{\mathop{\rm Re}\nolimits}

\makeatletter
\def\gsim{\compoundrel>\over\sim}
\def\lsim{\compoundrel<\over\sim}
\def\compoundrel#1\over#2{\mathpalette\compoundreL{{#1}\over{#2}}}
\def\compoundreL#1#2{\compoundREL#1#2}
\def\compoundREL#1#2\over#3{\mathrel
         {\vcenter{\hbox{$\m@th\buildrel{#1#2}\over{#1#3}$}}}}
\makeatother

\begin{document}
\newcommand{\Lagr}{\mathcal{L}}
\title{\vspace{-2cm} 
	\begin{flushright}
		{\normalsize INR-TH-2020-011}
	\end{flushright}
	\vspace{0.5cm} Sgoldstino signature in $hh$, $W^+W^-$ and $ZZ$ spectra at the LHC}
\author{S. Demidov${}^{1,2}$, D. Gorbunov${}^{1,2}$, E. Kriukova${}^{1,3}$\\
${}^1$ 
  Institute for Nuclear Research of the Russian Academy of Sciences, \\
Moscow 117312, Russia\\
${}^2$ Moscow Institute of Physics and Technology, \\
Dolgoprudny 141700, Russia\\ 
${}^3$ Lomonosov Moscow State University,\\
Moscow 119991, Russia\\
}

\date{}

\maketitle

\begin{abstract} 
In a supersymmetric extension of the Standard Model of particle
physics (SM) with low scale of supersymmetry breaking, sgoldstino of 
(sub)TeV mass can be observed at the Large Hadron Collider (LHC) as a 
peak in diboson mass spectra. Moreover, as a singlet with respect to
the SM gauge group, scalar sgoldstino can mix with the SM-like Higgs
boson and interfere in all neutral channels providing with the
promising signatures of  new physics if superpartners are heavy.
Sgoldstino couplings to the SM particles are determined by the
pattern of soft supersymmetry breaking parameters. Here we
concentrate on the cases with a noticeable sgoldstino contribution to
di-Higgs channel. Having found a
  phenomenologically viable region in the model parameter space 
where scalar sgoldstino, produced at the LHC in gluon fusion, decays
into a pair of the lightest Higgs boson  we give
predictions for 
corresponding cross section. Using the results of the LHC searches for
scalar resonances we place bounds on the supersymmetry breaking scale
$\sqrt{F}$ in this region of parameter space.  Remarkably, in this
region sgoldstino may also be observed in $W^+W^-$ and $ZZ$ channels,
yielding independent signatures, since their signal
strengths are related to  that of di-Higgs
  channel.   
\end{abstract}

%%%%%%%%%%%%%%%%%%%%%%%%%%%%%%%%%%%%%%%%%%%%%%%%%%%%%%%%%%%%%%%%%%%%%%%%%%%%%%%%%%%
\section{Introduction}

The first stages of LHC running culminated in 2012 with the discovery
of the Higgs boson~\cite{Aad:2012tfa,Chatrchyan:2012xdj}. Further
experimental program of the LHC experiments includes measurement and
precise determination of the Higgs boson coupling constants using data
on the Higgs boson production cross sections and decay widths.
Numerous models of new physics predict deviations of these coupling
constants from their SM values. In this respect, the Higgs boson
self-coupling is of great interest as several classes of SM extensions
imply some modification of the Higgs sector. Double Higgs  boson
production is one of the most sensitive observables to the Higgs boson
self-interaction and specific modifications of the Higgs sector. This
process is expected to be seen  with various signatures at LHC  (see
Ref.~\cite{DiMicco:2019ngk} for a review) operating in the
high luminosity regime (HL-LHC). A deviation of the cross section from the
SM prediction could indicate possible ways to extend the SM.

The cross section of the di-Higgs production in proton collisions in the
SM was calculated in \cite{Glover:1987nx, Plehn:1996wb, Dawson:1998py,
  Borowka:2016ypz, Grazzini:2018bsd, Chen:2019lzz}.
 The double Higgs boson production can be resonantly enhanced if the Higgs
sector contains new scalar(s) with the mass around TeV scale. Examples
of such scenarios were discussed in
Refs.~\cite{No:2013wsa,Chen:2014ask,Nakamura:2017irk,Lewis:2017dme,Grober:2017gut,Dawson:2017jja,Basler:2018dac}. In this paper we consider supersymmetric models with low scale
supersymmetry breaking (see e.g.~\cite{Ellis:1984kd,Ellis:1984xe,Brignole:1996fn, Gherghetta:2000qt, Gherghetta:2000kr, Brignole:2003cm, Navarro:2003he, Antoniadis:2011xi, Dudas:2012fa}). In this class of models the lightest
supersymmetric particle is gravitino which acquires its mass via
the super-Higgs mechanism~\cite{Volkov:1973jd, Deser:1977uq, Cremmer:1978iv}. The longitudinal component of gravitino
appears after spontaneous supersymmetry breaking as a derivative of
the Goldstone fermion -- 
goldstino $G$. The latter can be the fermionic component of a chiral
supermultiplet, 
\begin{equation}
  \label{Phi}
\Phi = \phi + \sqrt{2}\theta G + \theta^2F_\phi\,,
  \end{equation} 
where $\theta$ is the Grassmannian coordinate, $\phi$ is the complex scalar field called sgoldstino and
$F_\phi$ is the auxiliary field which acquires non-zero vev,
i.e. $\langle F_\phi\rangle = F$, triggering spontaneous SUSY
breaking in the entire model. Here we consider scenarios in which the scale $\sqrt{F}$ of
SUSY breaking is not very far from the electroweak scale while all the 
fields in the hidden sector except for those belonging to the goldstino multiplet are
heavy. The purpose of our study is to investigate possibility of the
resonant enhancement of di-Higgs production at the LHC due to
contribution of relatively light scalar sgoldstino. To make the picture as
clear as possible, yet realistic and phenomenologically viable, we
restrict ourselves to the case in which all superpartners of the SM
particles are also relatively heavy.

Though model-dependent, but typically the main sgoldstino production
mechanism in $pp$ collisions is through gluon-gluon
fusion~\cite{Perazzi:2000ty}. Being even with respect to $R$-parity,  
sgoldstino can decay into
pairs of the SM particles. We single out the part of the parameter space where
branching fraction of sgoldstino decay into pair of the Higgs bosons
is close to the largest possible value which is $25$\%. This region of the
model parameter space corresponds to sufficiently large mixing between
sgoldstino and the lightest Higgs boson which can result in an
amplification of the corresponding decay width and consequently di-Higgs production cross
section. In this regime sgoldstino decays dominantly into $hh$,
$W^+W^-$ and $ZZ$ with partial widths related as 1:2:1
respectively. We discuss the sgoldstino signature in spectra of these
final states at the LHC and calculate cross sections of 
the  corresponding processes. Using present experimental constraints on
production of the scalar resonances decaying into the heavy bosons we
obtain bounds on the SUSY breaking scale. 

This paper is organized as follows. In Section~\ref{lagr} we
describe the model Lagrangian and derive the potential of
scalar fields. We discuss interaction of sgoldstino and Higgs bosons
and mixing in this sector. We obtain sgoldstino-Higgs bosons trilinear
couplings. The scalar sgoldstino production channels and decay modes
are discussed in Section~\ref{production-and-decay}. There we outline
a region in the model 
parameter space where sgoldstino can decay into pair of the lightest
Higgs bosons as well as into the pair of the massive vector bosons
with a considerable probabilities. The resonant behavior of the
di-Higgs, $W^+W^-$ and $ZZ$ cross sections and their dependence 
on model parameters are investigated in Section~\ref{higgsprod}. There we
also present bounds on the supersymmetry breaking scale extracted from results
of ATLAS and CMS experiments in searches for di-boson
resonances. Our  findings are summarized in Section~\ref{conclude}.  
 Appendicies are reserved for explicit expressions for trilinear
  coupling constants of the neutral scalars.

%%%%%%%%%%%%%%%%%%%%%%%%%%%%%%%%%%%%%%%%%%%%%%%%%%%%%%%%%%%%%%%%%%%%%%%%%%%%%%%
\section{The model} \label{lagr}
We consider a supersymmetric extension of the Standard Model with
the chiral goldstino superfield \eqref{Phi}.
The model includes the Minimal Supersymmetric
Standard Model (MSSM) in which soft supersymmetry breaking terms are generated through
interactions with $\Phi$, when the auxiliary component of $F_\Phi$ gets
replaced with its non-zero vacuum expectation value, i.e. $\langle F_\phi\rangle
= F$. The Lagrangian
of the model can be written (see e.g.~\cite{Gorbunov:2001pd}) as the
sum 
\begin{equation}
  \label{eq:1}
  \Lagr=\Lagr_{K}+\Lagr_{W}+\Lagr_{gauge}+\Lagr_{\Phi}
\end{equation}
of the contributions from the K\"{a}hler potential $\Lagr_{K}$,
superpotential $\Lagr_{W}$, vector fields $\Lagr_{gauge}$
and a part which describes dynamics of the goldstino supermultiplet $\Lagr_{\Phi}$. 
The first term in~\eqref{eq:1} has the form
\begin{equation}
  \Lagr_{K}=\int d^2\theta d^2\bar{\theta}\sum_{k}\left(1-
  \frac{m_k^2}{F^2}\Phi^\dagger\Phi \right) \Phi_k^\dagger e^{g_1 V_1
    + g_2 V_2 + g_3 V_3}\Phi_k, 
\end{equation}
where the sum is taken over all matter superfields $\Phi_k$. The parts
$\Lagr_{W}$ and $\Lagr_{gauge}$ read 
\begin{multline}
  \Lagr_{W}=\int d^2\theta\epsilon_{ij}\left(
  \left(\mu-\frac{B}{F}\Phi\right) H_D^i
  H_U^j+\left(Y_{ab}^L+\frac{A^{L}_{ab}}{F}\Phi\right) L_a^j E_b^c
  H_D^i+\right.\\ \left.+\left(
  Y_{ab}^D+\frac{A^{D}_{ab}}{F}\Phi\right) Q_a^j D_b^c H_D^i+\left(
  Y_{ab}^U+\frac{A^{U}_{ab}}{F}\Phi\right) Q_a^iU_b^cH_U^j
  \right)+h.c.,  
\end{multline}
and
\begin{equation}
  \Lagr_{gauge}=\frac{1}{4}\sum_{a}\int
  d^2\theta\left(1+\frac{2M_a}{F}\Phi \right)
  \textnormal{Tr}W_{\alpha}W^{\alpha} + h.c.\,, 
\end{equation}
where sum goes over $a=3,2,1$, correspondingly to all the SM gauge
groups $SU(3)_c$, $SU(2)_w$, $U(1)_Y$.  
Here $\mu$ is the higgsino mixing parameter, $L, E$ are the left and
right lepton superfields, $Q, U, D$ are the superfields of left,
right up and right down quarks, respectively, $H_U, H_D$ are two Higgs 
doublet superfields, $Y^{L, D, U}$ are the matrices of the Yukawa
coupling constants. 
The interactions of $\Phi$ with other matter fields of MSSM, which
yield soft terms after SUSY breaking with the parameters $m_{k}^2$,
$M_{a}$, $B$, $A_{ab}^{L,D,U}$, are suppressed by $F$. The dynamics of
goldstino supermultiplet is 
modeled by the Lagrangian 
\begin{equation}
  \Lagr_\Phi=\int d^2\theta d^2\bar{\theta} \left(
  \Phi^{\dagger}\Phi-\frac{m_s^2+m_p^2}{8F^2}
  (\Phi^{\dagger}\Phi)^2-\frac{m_s^2-m_p^2}{12F^2}(\Phi^{\dagger}\Phi^3
  + \Phi^{\dagger 3}\Phi)\right) -\left( \int d^2\theta F\Phi +
  h.c. \right),   
\end{equation}
where $m^2_s$ and $m^2_p$ are mass parameters for
scalar and pseudoscalar sgoldstino fields. The model described above
should be considered as an effective theory which contains higher
order interaction terms suppressed by higher powers of $F$ and for
consistency we require smallness of the soft parameters, $m_{soft}\ll
\sqrt{F}$.  

\par Using the Lagrangian~\eqref{eq:1} one can obtain the scalar
potential of the Higgs and sgoldstino sector 
(see also \cite{Dyakonov2018})
\begin{equation}
  V=V_{11}+V_{12}+V_{21}+V_{22},
\end{equation}
\begin{equation} \label{eq:V11}
  V_{11}=\frac{g_1^2}{8}
  \left(1+\frac{M_1}{F}(\phi+\phi^*)\right)^{-1}\left[h_d^\dagger
    h_d-h_u^\dagger h_u-\frac{\phi^*\phi}{F^2}\left(m_d^2h_d^\dagger
    h_d-m_u^2h_u^\dagger h_u\right)\right]^2, 
\end{equation}
\begin{multline} \label{eq:V12}
  V_{12}=\frac{g_2^2}{8}
  \left(1+\frac{M_2}{F}(\phi+\phi^*)\right)^{-1}\left[h_d^\dagger
    \sigma_a h_d+h_u^\dagger \sigma_a
    h_u-\frac{\phi^*\phi}{F^2}\left(m_d^2h_d^\dagger \sigma_a
    h_d+m_u^2h_u^\dagger \sigma_a h_u\right)\right]^2. 
\end{multline}
\begin{multline} \label{eq:V21}
  V_{21}=\left(1-\frac{m_s^2+m_p^2}{2F^2}\phi^*\phi-\frac{m_s^2-m_p^2}{4F^2}\left(
  \phi^2+\phi^{*2}\right)-\frac{m_u^2}{F^2}h_u^\dagger
  h_u-\frac{m_d^2}{F^2}h_d^\dagger h_d-\frac{m_u^4}{F^4}\phi^*\phi
  h_u^\dagger h_u-\right. \\ \left. -\frac{m_d^4}{F^4}\phi^*\phi
  h_d^\dagger h_d\right)^{-1}  \abs{F+\left(-h_d^0
    h_u^0+H^-H^+\right)\left(\frac{B}{F}-\frac{m_u^2+m_d^2}{F^2}
    \phi^*\left(\mu-\frac{B}{F}\phi\right)\right)}^2,   
\end{multline}
\begin{equation} \label{eq:V22}
  V_{22}=\frac{\mu^2}{F^2}\abs{\phi}^2\left(m_u^2h_d^\dagger
  h_d+m_d^2h_u^\dagger h_u\right)+\abs{\mu-\frac{B}{F}\phi}^2
  \left(h_d^\dagger h_d+h_u^\dagger h_u\right). 
\end{equation}
Here $h_d = \left(h_d^0, H^-\right)^T$, $h_u=\left(H^+, h_u^0\right)^T$ are the
Higgs doublets and we consider pall the parameters entering the
potential to be real.
As $\sqrt{F}$ is the 
largest scale in the model, we are interested in the scalar potential
to the leading order in $1/F$. Moreover, since each sgoldstino field
enters the interaction terms with factor $1/F$, rates of multi-sgoldstino
processes are naturally parametrically suppressed. Therefore, the most promising
processes are those involving single sgoldstino production.

Now we expand the scalar fields around their vacuum expectation values.
For the Higgs fields we have, as usual~\cite{Martin:1997ns}  
\begin{gather}
  \label{eq:hu}
  h_u^0=v_u+\frac{1}{\sqrt{2}}(h\cos{\alpha}+H\sin{\alpha})
  +\frac{i}{\sqrt{2}}A\cos{\beta}, \\ 
\label{eq:hd}
  h_d^0=v_d+\frac{1}{\sqrt{2}}(-h\sin{\alpha}+H\cos{\alpha})
  +\frac{i}{\sqrt{2}}A\sin{\beta}\,, 
\end{gather}
where $v_u\equiv v\sin{\beta}$, $v_d\equiv v\cos{\beta}$, $v=174$
GeV. Sgoldstino field is expanded as 
\begin{equation} \label{eq:sgld}
  \phi=\frac{1}{\sqrt{2}}(s+ip)\,,
\end{equation}
where $s$ and $p$ are its scalar and pseudoscalar components. In
general, the sgoldstino field acquires a non-zero vev
$\langle\phi\rangle = v_\phi$ and one can show \cite{Petersson:2011in} that $v_\phi$
scales as $1/F$ for scalar sgoldstino heavier than the lightest Higgs
boson. Therefore, the terms with sgoldstino vev 
would contribute to the terms with smaller number of sgoldstino fields
but suppressed by powers of $1/F^2$ and hence are 
neglected in what follows. 
Naturally, for consistency reasons, we set $v_\phi=0$. Likewise, 
point $h^0_u=v_u$, $h^0_d=v_d$ is a true vacuum up to corrections
suppressed by powers of $1/F^2$, which we ignore\footnote{Let us note
  that for $\sqrt{F}\sim$ few TeV the corrections of order $1/F^2$ to
  the scalar potential can be valuable and contribute considerably to
the mass of the lightest Higgs boson~\cite{Antoniadis:2014eta}.  For
the choice of parameters 
considered in the present study these corrections are negligible.}. 

One can substitute the expansions \eqref{eq:hu}--\eqref{eq:sgld} into the potential
\eqref{eq:V11}--\eqref{eq:V22} and find the sgoldstino-Higgs boson
vertices to the leading order in $1/F$. In order to simplify
expressions for the interaction terms, let us introduce the notations 
\begin{equation} \label{eq:def}
  m_Z^2\equiv\frac{g_1^2+g_2^2}{2}v^2, \hspace{1 cm} m_A^2\equiv
  m_u^2+m_d^2+2\mu^2\,,
\end{equation}
which correspond to tree-level squared masses of $Z$-boson and Higgs
pseudoscalar $A$. For the true vacuum there are 
no terms linear in $h, H$ in the potential, and so no linear terms
within our approximation. Minimization of the
scalar potential results in the the following relationships
\cite{Martin:1997ns}   
\begin{equation}
  \frac{1}{2}m_Z^2+\mu^2=\frac{m_d^2-m_u^2\tan^2{\beta}}{\tan^2{\beta}-1},
\end{equation} 
\begin{equation} \label{eq:B}
  \sin{2\beta}=\frac{2B}{m_A^2}.
\end{equation}
The mixing angle $\alpha$ in Eqs.~\eqref{eq:hu} and~\eqref{eq:hd} is
chosen in the standard way~\cite{Martin:1997ns}
\begin{equation} \label{eq:tan}
  \frac{\tan{2\alpha}}{\tan{2\beta}}=\frac{m_A^2+m_Z^2}{m_A^2-m_Z^2}.
\end{equation}
Once again, any possible corrections suppressed by powers of $1/F^2$
are small and neglected here. At the same time the loop corrections to
the effective potential should be large enough to generate observed
value of the Higgs boson mass. Their size is determined mainly by
interactions with top-stop sector. In what follows we do not discuss
squark sector of the model and just assume squarks to be sufficiently
heavy. Therefore, we treat the quantum corrections which
contribute to the Higgs boson masses as obligatory and of the  
size sufficient for saturating the lightest Higgs boson mass. Consequently, we neglect
radiative corrections to the mixing term between $h$ and $H$, that is
justified by the decoupling regime we assume in this study.

The mass terms in the Lagrangian for the scalar $(H,h,s)$ and
pseudoscalar $(A,p)$ fields have the following matrix form 
\begin{equation}
  \label{eq:mmatrix}
  \frac{1}{2}
  \begin{pmatrix}
    H&h&s\\
  \end{pmatrix}
  \begin{pmatrix} 
    m_H^2 & 0 & Y/F \\
    0 & m_h^2 & X/F \\
    Y/F & X/F & m_s^2 \\
  \end{pmatrix}
  \begin{pmatrix}
    H\\h\\s\\
  \end{pmatrix}+\frac{1}{2}
  \begin{pmatrix}
    A&p\\
  \end{pmatrix}
  \begin{pmatrix}
    m_A^2 & Z/F\\
    Z/F & m_p^2\\
  \end{pmatrix}
  \begin{pmatrix}
    A\\p\\
  \end{pmatrix}, 
\end{equation}
where the mixing terms read
\begin{multline} \label{eq:X}
  X\equiv v\biggl(\frac{g_1^2M_1+g_2^2M_2}{2}v^2\cos{2\beta}\sin{(\alpha+\beta)}+ \\
  +\mu
  m_A^2\sin{2\beta}\sin{(\alpha-\beta)}+\left(m_A^2-2\mu^2\right)\mu\cos{(\alpha+\beta)}
  \biggr),    
\end{multline}
\begin{multline} \label{eq:Y}
  Y\equiv -v\biggl(\frac{g_1^2M_1+g_2^2M_2}{2}v^2\cos{2\beta}\cos{(\alpha+\beta)}+
  \\+\mu
  m_A^2\sin{2\beta}\cos{(\alpha-\beta)}+\left(2\mu^2-m_A^2\right)\mu\sin{(\alpha+\beta)}
  \biggr), 
\end{multline}
\begin{equation} \label{eq:Z}
  Z\equiv \mu v\left(m_A^2-2\mu^2\right).
\end{equation}
According to the above discussion the diagonal values of the mass
matrices, i.e. $m_h^2$, $m_H^2$ and $m_A^2$ already contain leading 
quantum corrections. 
Recall, that our primary goal is to consider the possible
sgoldstino contribution to the pair productions of
the lightest Higgs bosons $h$ which are interpreted as the scalar
Higgs-like particles discovered at the LHC. We present below expressions
for the relevant trilinear coupling constants in the scalar
sector. Corresponding part of the interaction Lagrangian is
\be
\label{eq:trilinear}
\Lagr_{trilinear} =
C_{hhh}h^3+C_{hhH}h^2H+C_{hHH}hH^2+C_{HHH}H^3+C_{shh}sh^2+C_{sHH}sH^2\,,
\ee
where
\begin{gather}
  C_{hhh} \equiv
  \frac{1}{2\sqrt{2}}\frac{m_Z^2}{v}\cos{2\alpha}\sin{(\alpha+\beta)},
  \\ 
  C_{HHH} \equiv \frac{1}{2\sqrt{2}}\frac{m_Z^2}{v}
  \cos{2\alpha}\cos{(\alpha+\beta)},\\ 
\label{eq:hhH}
C_{hhH} \equiv  \frac{1}{4\sqrt{2}}\frac{m_Z^2}{v}
\left(\cos{(\alpha-\beta)-3\cos{(3\alpha+\beta)}} \right), \\ 
\label{eq:hHH}
C_{hHH} \equiv -\frac{1}{4\sqrt{2}}\frac{m_Z^2}{v}
\left(\sin{(\alpha-\beta)+3\sin{(3\alpha+\beta)}} \right) 
\end{gather}
are the MSSM Higgs trilinear couplings and
\begin{multline} \label{eq:oldsHH}
C_{sHH} \equiv \frac{1}{F\sqrt{2}}\left(-\frac{v^2}{4}
\left(g_1^2M_1+g_2^2M_2\right)\left(2\cos{2\alpha}\cos{2\beta}
-\sin{2\alpha}\sin{2\beta}+1\right)+\right. \\  \left.
+\mu\sin{2\alpha}\left[\frac{m_A^2}{2}\left(1-\frac{\sin{2\beta}}
  {\sin{2\alpha}}\right)-\mu^2\right]\right).    
\end{multline}
\begin{multline} \label{eq:oldshh}
C_{shh} \equiv \frac{1}{F\sqrt{2}}\left(\frac{v^2}{4}
\left(g_1^2M_1+g_2^2M_2\right)\left(2\cos{2\alpha}\cos{2\beta}
-\sin{2\alpha}\sin{2\beta}-1\right)
-\right. \\ \left. -\mu\sin{2\alpha} \left[\frac{m_A^2}{2}
  \left(1+\frac{\sin{2\beta}}{\sin{2\alpha}}\right)
  -\mu^2\right]\right) 
\end{multline}
are the trilinear coupling constants of the interaction between scalar
sgoldstino and the Higgs bosons. Other trilinear couplings in the
scalar sector are not relevant for 
the di-Higgs boson production in the model to the leading order in  
$1/F$. However, in Appendix \ref{trilincoef} we present
expressions for those interaction terms containing single sgoldstino
and  
two Higgs bosons for completeness.  They agree with the similar
expressions in Refs.\,\cite{Asano:2017fcp,Dyakonov2018}.  
		
At $F\rightarrow\infty$ the decompositions~\eqref{eq:hu},
\eqref{eq:hd} and~\eqref{eq:sgld} make the mass matrices of the
component fields $(H,h,s)$ and $(A,p)$ diagonal at tree level.
At finite $F$ and relatively light sgoldstino considered in this study 
the mixing between these fields can result in a considerable
modification of the model phenomenology. Being interested primary in
$hh$ final state we concentrate below on the mixing in the scalar
sector, i.e. between $H$, $h$ and $s$, and
denote the mass states as $\tilde{H}$, $\tilde{h}$ and $\tilde{s}$.
Similar transformation can be also performed for
the mass matrix of pseudoscalars $A$, $p$ by introducing fields $\tilde{A},
\tilde{p}$.  
	
In order to make the $3\times 3$ mass squared matrix in
Eq.~\eqref{eq:mmatrix} diagonal, we consider the following rotation which
is parameterized by the mixing angles $\phi$, $\psi$, $\theta$ 
\begin{multline} \label{eq:matr}
  \begin{pmatrix}
    \cos{\phi}&\sin{\phi}&0\\
    -\sin{\phi}&\cos{\phi}&0\\
    0&0&1\\
  \end{pmatrix}
  \begin{pmatrix}
    \cos{\psi}&0&\sin{\psi}\\
    0&1&0\\
    -\sin{\psi}&0&\cos{\psi}\\
  \end{pmatrix}
  \begin{pmatrix}
    1&0&0\\
    0&\cos{\theta}&\sin{\theta}\\
    0&-\sin{\theta}&\cos{\theta}\\
  \end{pmatrix} \times\\\times
  \begin{pmatrix} 
    m_H^2 & 0 & Y/F \\
    0 & m_h^2 & X/F \\
    Y/F & X/F & m_s^2 \\
  \end{pmatrix} \times\\\times
  \begin{pmatrix}
    1&0&0\\
    0&\cos{\theta}&-\sin{\theta}\\
    0&\sin{\theta}&\cos{\theta}\\
  \end{pmatrix}
  \begin{pmatrix}
    \cos{\psi}&0&-\sin{\psi}\\
    0&1&0\\
    \sin{\psi}&0&\cos{\psi}\\
  \end{pmatrix}
  \begin{pmatrix}
    \cos{\phi}&-\sin{\phi}&0\\
    \sin{\phi}&\cos{\phi}&0\\
    0&0&1\\
  \end{pmatrix}.
\end{multline}
Hereinafter we work in the approximation of the small mixing angles. Given
that the mixing terms are suppressed by $1/F$ we take the angles
$\phi$, $\psi$, $\theta$ to the leading non-zero order in $1/F$. Then
by writing the condition of zero off-diagonal elements in the
first-order approximation in \eqref{eq:matr} one can obtain the
expressions for the mixing angles 
\begin{gather} \label{eq:psi}
  \psi=\frac{Y}{F(m_H^2-m_s^2)}\,, \\
\label{eq:theta}
  \theta=\frac{X}{F(m_h^2-m_s^2)}\,,
\end{gather}
while as 1-2 and 2-1 elements in the mass matrix~\eqref{eq:mmatrix} are
equal to zero, the third mixing angle $\phi$ appears to be of 
the second order in $1/F$:  
\begin{equation} \label{eq:phi}
  \phi=\frac{XY}{F^2(m_h^2-m_s^2)(m_H^2-m_h^2)}\,.
\end{equation}
It can be easily seen from the equations \eqref{eq:psi},
\eqref{eq:theta}, \eqref{eq:phi} that the mixing angles $\psi$, $\theta$,
$\phi$ can be quite large if the corresponding masses $m_h$, $m_H$, $m_s$ are close
to each other. In what follows we do not consider the case of 
degenerate scalars to avoid large values of the mixing angles (that
pattern may be unrealistic in any cases given measurements at the LHC).   
	
Rotation to the mass eigenstates denoted as $\tilde{H}$, $\tilde{h}$, 
$\tilde{s}$ is given by $\begin{pmatrix} 
  H & h&s\\ 	\end{pmatrix}=\begin{pmatrix}
\tilde{H}&\tilde{h}&\tilde{s}\\ \end{pmatrix}C	$, where  
in the approximation of small mixing the rotation matrix $C$ can be
written as 
\begin{equation}
  \label{eq:mix_matr}
  C=\begin{pmatrix}
  1-\psi^2/2&\phi-\psi\theta&\psi\\
  -\phi&1-\theta^2/2&\theta\\
  -\psi&-\theta&1-\psi^2/2-\theta^2/2
  \end{pmatrix}.
\end{equation}
Similar calculations can be performed for the mass matrix of the
pseudoscalar fields. Corresponding mixing angle $\xi$ reads 
\begin{equation} \label{eq:xi}
  \xi=\frac{Z}{F(m_A^2-m_p^2)}.
\end{equation}
In the basis of mass eigenstates the trilinear coupling constants
which are relevant for resonant production of the pair of the lightest
Higgs bosons can be written as follows  
\begin{equation} 
\label{eq:shh}
 {\cal L}_{\tilde{s}\tilde{h}\tilde{h}} =
 \left(C_{shh}-3C_{hhh}\theta-C_{hhH}\psi\right)\tilde{s}\tilde{h}\tilde{h}
 \equiv C_{\tilde{s}\tilde{h}\tilde{h}}\tilde{s}\tilde{h}\tilde{h}\,, 
\end{equation}
where we omitted the terms of order $1/F^2$ and higher. 

In the next Sections these vertices are used in the calculation of
sgoldstino decay rate into a pair of neutral Higgs bosons and of the
resonant double-Higgs production cross section.
Expressions for other vertices with one sgoldstino and
two Higgs bosons ($\tilde{s}\tilde{h}\tilde{H}$,
$\tilde{s}\tilde{A}\tilde{A}$, $\tilde{s}\tilde{H}^{+}\tilde{H}^{-}$,
$\tilde{p}\tilde{A}\tilde{H}$, $\tilde{p}\tilde{A}\tilde{h}$), 
irrelevant for the present study, can be
found in Appendix~\ref{trilincoefnew}. 

In the next sections we will discuss production and decay of scalar
sgoldstino. The relevant part of the Lagrangian describing interactions of
the scalar sgoldstino $s$ with SM vector bosons has the form 
\begin{eqnarray} 
\label{eq:sg_lagr}
{\cal L}^{eff}_{s} & = &
-\frac{M_{2}}{\sqrt{2}F}sW^{\mu\nu *}W_{\mu\nu}
-\frac{M_{ZZ}}{2\sqrt{2}F}sZ^{\mu\nu}Z_{\mu\nu}
-\frac{M_{Z\gamma}}{\sqrt{2}F}sF^{\mu\nu}Z_{\mu\nu} \\
& & -\frac{M_{\gamma\gamma}}{2\sqrt{2}F}sF^{\mu\nu}F_{\mu\nu}
 - \frac{M_3}{2\sqrt{2}F}s~G^{a\mu\nu}G^{a}_{\mu\nu}\,, \label{eq:sg_glue}
\end{eqnarray}
where $W_{\mu\nu}$, $Z_{\mu\nu}$, $F_{\mu\nu}$ and $G^a_{\mu\nu}$
are the field strength tensors for the $W^\pm$-bosons, $Z$-boson,
photon and gluons respectively; the mass parameters in front of
the couplings read 
\begin{gather}
M_{ZZ} \equiv M_{1}\sin^{2}{\theta_{W}} +
M_{2}\cos^{2}{\theta_{W}},\;\;\;
M_{\gamma\gamma} \equiv M_{1}\cos^{2}{\theta_{W}} +
M_{2}\sin^{2}{\theta_{W}}\,,\\
M_{Z\gamma} \equiv (M_2-M_1)\cos{\theta_W}\sin{\theta_W}\,.
\end{gather}
Interactions~\eqref{eq:sg_lagr} along with~\eqref{eq:trilinear} and
sgoldstino-Higgs mixing given by~\eqref{eq:mix_matr} determine
phenomenology of TeV scale scalar sgoldstino at hadron colliders. Results of
our analysis depend upon the following parameters: 
$\tan{\beta}$, $\mu$, $m_A$, $m_s$, $M_1$, $M_2$, $M_3$, $F$. 
Since formulas of Section~\ref{lagr} are 
obtained in the small mixing angles approximation, one should keep
only those models in the parameter space that correspond to small mixing
angles $\psi$, $\theta$ (for the numerical calculations we select the
models with $\theta<0.3$, $\psi<0.3$).

%%%%%%%%%%%%%%%%%%%%%%%%%%%%%%%%%%%%%%%%%%%%%%%%%%%%%%%%%%%%%%%%%%%%%%%%%%
\section{Sgoldstino production and decay channels}
\label{production-and-decay}
To understand possible size of the resonant  double Higgs boson production
in the model we need to study sgoldstino production in proton
collisions and its subsequent decay. Here we refer to Refs.~\cite{Perazzi:2000id, Perazzi:2000ty, Asano:2017fcp, Astapov:2014mea, Demidov:2017rmi, Ding:2016udc, Gorbunov:2002er} for previous
studies of sgoldstino collider phenomenology. To simplify notations,
in what follows we use $s$ and $h$ for the mass eigenstates (which were
denoted as $\tilde s$ 
and $\tilde h$ in the previous Section) and take into account
contributions from the sgoldstino-Higgs mixing to the leading order in
$1/F$. 

\subsection{Production of scalar sgoldstino}
\label{production}
\begin{figure}
  \begin{center}
    \begin{subfigure}{0.3\textwidth}
      \includegraphics[width=0.8\textwidth]{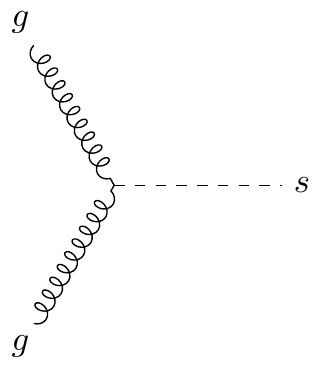}
      \caption{}
      \label{fig:ggs}
    \end{subfigure}
    \begin{subfigure}{0.5\textwidth}
      \includegraphics[width=\textwidth]{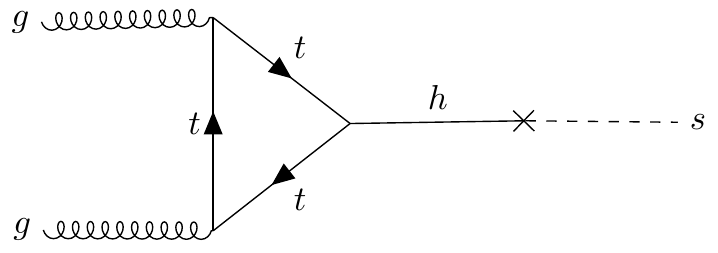}
      \caption{}
      \label{fig:gghs}
    \end{subfigure}
    \caption{Feynman diagrams for the sgoldstino production
      via gluon fusion to the leading order in $1/F$; the cross refers
      to the Higgs-sgoldstino mixing.}			
  \end{center}
\end{figure}

For typical hierarchy between the soft SUSY breaking parameters the
dominant sgoldstino production mechanism is gluon
fusion~\cite{Perazzi:2000ty}. This process, $gg \rightarrow s$, occurs 
in this model already at the tree level due to sgoldstino interaction
with gluons governed by the coupling $M_3/F$, see
Eq.~\eqref{eq:sg_lagr}. Mixing with the Higgs bosons yields 
additional contributions to the amplitude of this process, see
Figs.~\ref{fig:ggs}, \ref{fig:gghs}. The contribution from
sgoldstino mixing with heavy neutral Higgs (similar to Fig. \ref{fig:gghs})
is suppressed by  $1/\tan\beta$ from the $Ht\bar t$ vertex 
and also by square of its mass $m_H^2$ in the expression for the
corresponding mixing angle, therefore we neglect it in the following
calculation of sgoldstino production cross section.  
The leading order cross section of the sgoldstino production in gluon fusion
can be calculated using the parton distribution function (PDF) 
$g(x, m_s^2)$ of gluons in proton as
\begin{equation} \label{eq:sgldprod}
  \sigma_{prod}(pp\rightarrow s)=\sigma_0\tau\int_{\tau}^{1}
  \frac{\dd{x}}{x} g\left(x, m_s^2\right)g\left(\frac{\tau}{x},
  m_s^2\right), 
\end{equation}
with
\begin{equation}
  \label{eq:sigma0}
  \sigma_0=\frac{\pi}{32}\abs{\frac{M_3}{F}+\frac{\alpha_s(m_s)\theta}{6\pi
      v}A\left(\frac{4m_t^2}{m_s^2}\right)}^2, \hspace{0.7cm}
  \tau=\frac{m_s^2}{S}.  
\end{equation}
Here $\sqrt{S}$ is the center-of-mass energy in $pp$ collisions and
$A\left(\frac{4m_t^2}{m_s^2}\right)$ is the loop factor (see,
e.g. \cite{Spira:1997dg})
\begin{equation*}
  A(\tau_q)=\frac{3}{2}\tau_q\left(1+\left(1-\tau_q\right)f(\tau_q)\right), 
\end{equation*}
with 
\begin{equation*}
  f(\tau_q)=\begin{cases}
  \arcsin^2\frac{1}{\sqrt{\tau_q}}, & \text{if } \tau_q\geq 1\\
  -\frac{1}{4}\left[\log\frac{1+\sqrt{1-\tau_q}}{1-\sqrt{1-\tau_q}}
    -i\pi\right] ^2, & \text{if } \tau_q < 1. 
  \end{cases}
\end{equation*}
Note that with the mixing angle $\theta$ given by~\eqref{eq:theta}
both amplitudes in~\eqref{eq:sigma0} are of the same order 
in $1/F$ and in the loop part of the amplitude we leave only the
dominant contribution with the heavy top quark loop. As we assume the
superpartners (in particular, squarks) to be relatively heavy, their
contribution to this process is negligible.

The NLO corrections to the production cross section for a scalar which
has both effective gluon interaction~\eqref{eq:sg_glue} and Yukawa
interaction with top quark was calculated
in~\cite{Anastasiou:2016hlm}.  According to these results,
corresponding $K$-factor, i.e. $K_p=\sigma_{NLO}/\sigma_{LO}$, is about
2 for chosen sgoldstino mass interval  and $\sqrt{S}=13,
14$~TeV. Since 
dependence of the $K$-factor on the relevant parameters in the
considered parameter space is weak, we use the value $K_p=2$ to correct
the leading order sgoldstino production cross section calculated using
CTEQ6 Parton Distribution Functions (the CTEQ6L1 PDF-set for computations in
the leading order)~\cite{Pumplin:2002vw}.

\begin{figure}
  \begin{center}
    \begin{subfigure}{0.48\textwidth}
      \includegraphics[width=\textwidth]{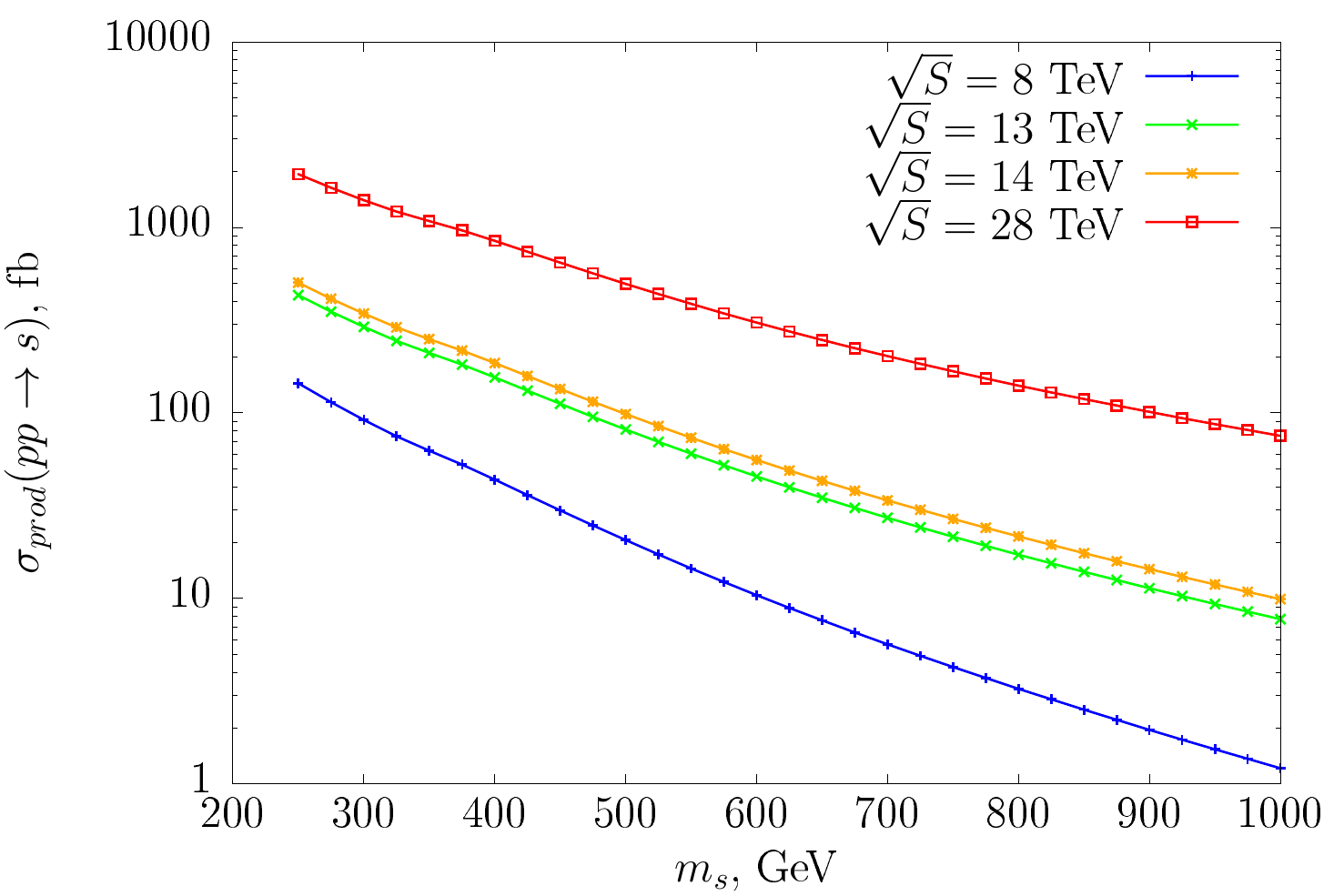}
      \caption{}
      \label{fig:prodsmalltheta}
    \end{subfigure}
    \begin{subfigure}{0.48\textwidth}
      \includegraphics[width=\textwidth]{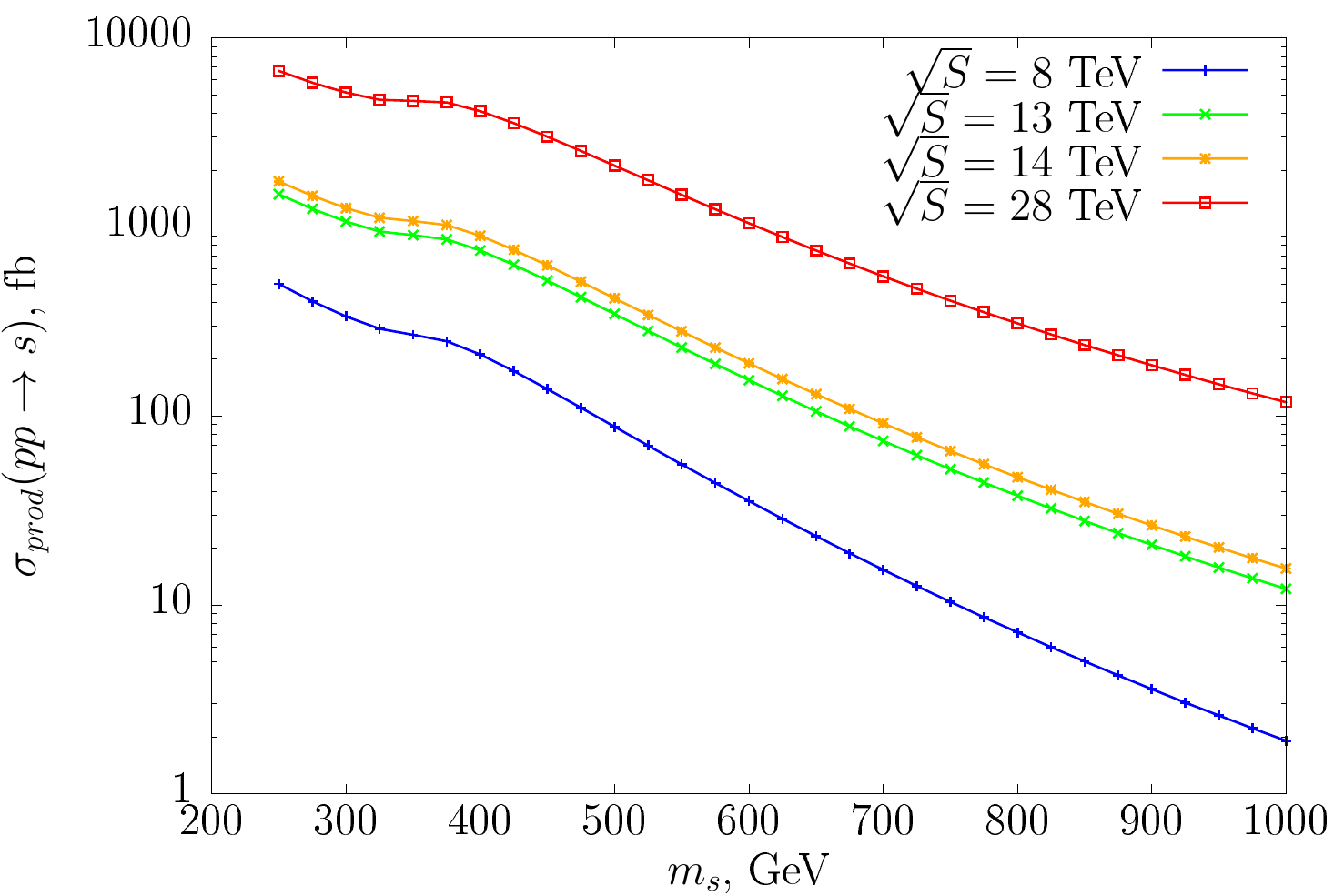}
      \caption{}
      \label{fig:prodbigtheta}
    \end{subfigure}
    \caption{Sgoldstino production cross section for a set of total
      energies of protons $\sqrt{S}$, and model parameters fixed as: $M_3=3$
      TeV, $\sqrt{F}=20$ TeV, $K_p=2$, (a) $\theta=0.02$, (b)
      $\theta=0.2$.}  
    \label{fig:prod}			
  \end{center}
\end{figure}

The scalar sgoldstino production cross section is shown in Fig.~\ref{fig:prod}
for the fixed values of parameters $M_3 = 3$ TeV, $\sqrt{F} =
20$ TeV, for 
center-of-mass energies $\sqrt{S} =$ 8, 13, 14 and 28~TeV as well as
for 
two values of the mixing angle $\theta=0.02$ (left panel (a)) and $0.2$
(right panel (b)). In the case of relatively small sgoldstino-Higgs mixing
the cross section of sgoldstino production presented in
Fig.~\ref{fig:prodsmalltheta} is dominated by the first term
in~\eqref{eq:sigma0}, see Fig.~\ref{fig:ggs}. In the opposite case of
large mixing angle, the cross section in Fig.~\ref{fig:prodbigtheta}
reveals a non-trivial behavior related to the sgoldstino mass
dependent contribution from the loop factor in the amplitude with
sgoldstino-Higgs mixing, cf. Fig.~\ref{fig:gghs}. For light
sgoldstinos, the contribution from the mixing term dominates in the
production cross section, see 
Fig.~\ref{fig:gghs}. The effect is reduced with increasing sgoldstino
mass and almost disappears as $m_s$ approaches 1\,TeV, where the
first term in~\eqref{eq:sigma0}, Fig.\,\ref{fig:ggs}, fully dominates.   

Other sgoldstino production channels~\cite{Perazzi:2000ty} include
vector boson fusion, sgoldstino associated production with a vector
boson and $t\bar{t}s$ production. They are suppressed as
compared to gluon fusion for the parameter space discussed in this
study (even the sgoldstino-Higgs mixing does not change the day). In
the considered region of sgoldstino mass $m_s$ the gluon fusion gives the
dominant contribution to production of virtual Higgs boson that mixes
with sgoldstino. Since we consider regime of  small values of mixing
angle $\theta$, the diagrams with sgoldstino-Higgs mixing are
accordingly suppressed as compared to one in Fig. \ref{fig:ggs}. Therefore we
consider for them only the gluon fusion channel of Higgs boson production.

\subsection{Sgoldstino decays}
\label{numcrsec}
\begin{figure}[t]
  \begin{center}
  	\begin{subfigure}{0.61\textwidth}
  		\includegraphics[width=0.25\textwidth]{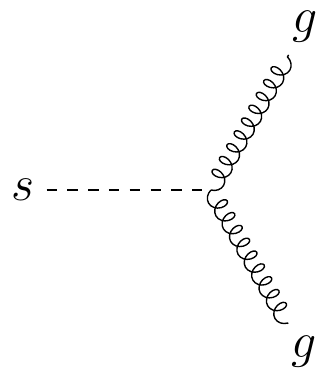}
  		\includegraphics[width=0.69\textwidth]{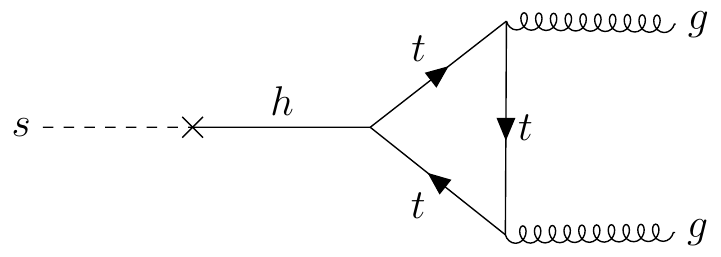}
  		\caption{}
  		\label{fig:sgg}
  	\end{subfigure}
    \begin{subfigure}{0.18\textwidth}
      \includegraphics[width=0.95\textwidth]{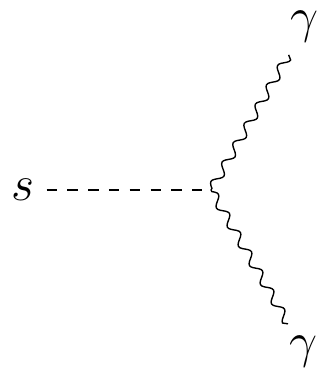}
      \caption{}
      \label{fig:sgaga}
    \end{subfigure}
    \begin{subfigure}{0.18\textwidth}
      \includegraphics[width=0.95\textwidth]{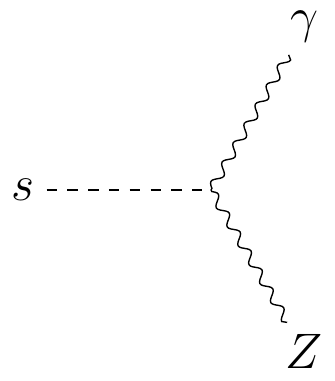}
      \caption{}
      \label{fig:sZga}
    \end{subfigure}
    \begin{subfigure}{0.59\textwidth}
    	\includegraphics[width=0.237\textwidth]{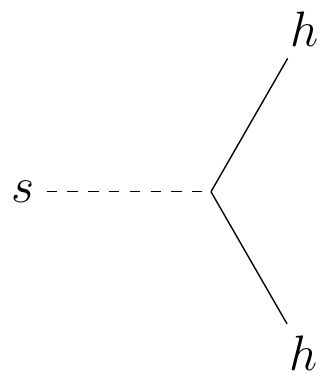}
    	\includegraphics[width=0.35\textwidth]{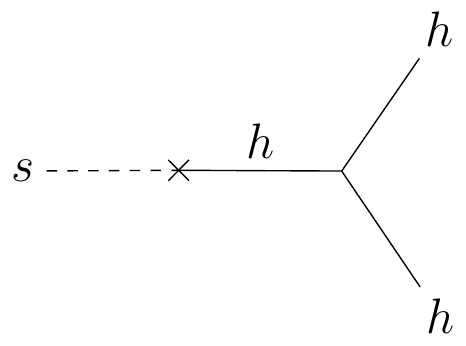}
    	\includegraphics[width=0.35\textwidth]{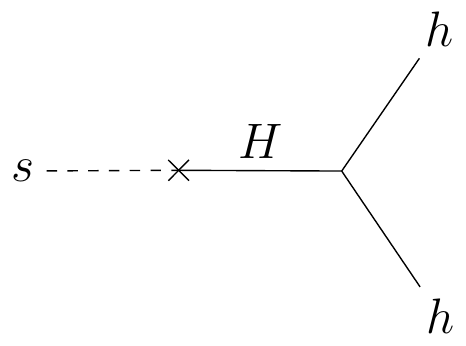}
    	\caption{}
    	\label{fig:shh}
    \end{subfigure}
    \begin{subfigure}{0.35\textwidth}
      \includegraphics[width=\textwidth]{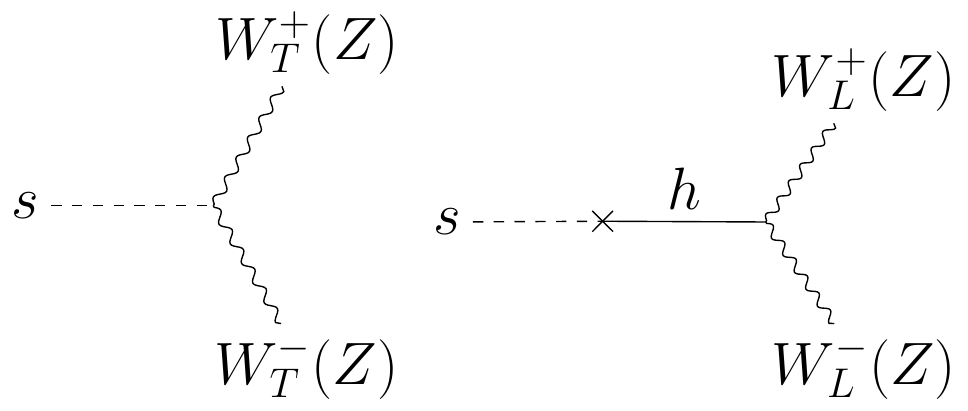}
      \caption{}
      \label{fig:sWW}
    \end{subfigure}
    \caption{\label{fig:sdecays} Feynman diagrams for main sgoldstino
      decay channels.}	 
  \end{center}
\end{figure}

Let us discuss the main decay modes of (sub)TeV sgoldstinos. The Feynman diagrams
for the dominant decay channels are shown in
Fig.~\ref{fig:sdecays}.
Corresponding amplitudes are determined by the
interactions~\eqref{eq:sg_lagr} and sgoldstino-Higgs mixing. The
latter is taken into account in the analysis to the leading order in
$1/F$.  

We start with the sgoldstino decay into a pair of gluons, 
$s\rightarrow gg$. As in the case of sgoldstino production in gluon
fusion there are two amplitudes in the leading order. Indeed,
sgoldstino can decay directly into gluons or
via mixing with the lightest Higgs boson, see Fig.~\ref{fig:sgg}.
Since in the leading order the decay to gluons is just the process
cross symmetric to the gluon fusion, we find in the formula for
sgoldstino rate to gluons the similar expression as in formula for
production cross section \eqref{eq:sigma0}, namely 
\begin{equation} \label{eq:sgg}
  \Gamma\left(s\rightarrow
  gg\right)=\frac{1}{4\pi}\abs{\frac{M_3}{F}
    +\frac{\alpha_s(m_s)\theta}{6\pi  
      v}A\left(\frac{4m_t^2}{m_s^2}\right)}^2 m_s^3.  
\end{equation}
In the absence of the mixing (i.e. when $\theta=0$) this expression
can be found in Ref.\,\cite{Perazzi:2000id}. We include NLO QCD
corrections by introducing the factor $K_d=1.6$ for gluon decay
channel of sgoldstino. In the considered regime of sgoldstino decaying 
dominantly into Higgs and massive vector bosons, its precise value is
not very important and we conservatively choose it close to that of
predicted for gluonic decay of the Higgs-like scalar (see Fig.~7
in~\cite{Spira:1997dg}).  

In the sgoldstino decays to either pair of photons $s\rightarrow 
\gamma \gamma$ (Fig. \ref{fig:sgaga}), or photon and Z-boson
$s\rightarrow \gamma Z$ (Fig. \ref{fig:sZga}), we can neglect the
mixing contribution, since the
Higgs boson $h$ decays into these particles through loops and
the corresponding amplitudes are highly suppressed by gauge and loop
factors. Therefore, 
the corresponding decay widths reads~\cite{Perazzi:2000id}  
\begin{gather}
  \Gamma\left(s\rightarrow \gamma
  \gamma\right)=\frac{1}{32\pi}\frac{1}{F^2}M_{\gamma\gamma}^2 m_s^3, \\
  \Gamma\left(s\rightarrow \gamma Z\right)=
  \frac{1}{16\pi}\frac{M_{Z\gamma}^2}{F^2}m_s^3\left(1-\frac{m_Z^2}{m_s^2}
  \right)^3. 
\end{gather}

Next we consider the scalar sgoldstino to be heavy enough for the
decay into a pair of the lightest Higgs bosons
(Fig. \ref{fig:shh}). The decay width can be
calculated with help of the trilinear coupling in Eq.~\eqref{eq:shh} as 
\begin{equation}
  \Gamma({s}\rightarrow
  {h}{h})=\frac{C^2_{\tilde{s}\tilde{h}\tilde{h}}}{8\pi
    m_s}\sqrt{1-\frac{4m^2_h}{m_s^2}}\,. 
\end{equation}
\par The case of sgoldstino decays to massive vector bosons is more 
involved~\cite{Perazzi:2000id}. Let us consider the decay of
sgoldstino into a pair of W-bosons (Fig.~\ref{fig:sWW}). The
sgoldstino Lagrangian contains two different terms responsible for
the interaction between sgoldstino and W-bosons. The first one is
the contribution of the first term in~\eqref{eq:sg_lagr} which is
determined by the coupling constant
\begin{equation} 
  C_{sWW_T}=-\frac{M_2}{F\sqrt{2}}.
\end{equation}   
This term originates from the sgoldstino interaction
Lagrangian~\eqref{eq:sg_lagr} and is not affected by mixing with the
neutral Higgs bosons (to the leading order in $1/F$). Another 
contribution is related to sgoldstino-Higgs mixing and the
interactions of the Higgs bosons with $W^\pm$. The corresponding
interaction Lagrangian has the form
\begin{equation}
  C_{{s}WW_L}m_W^2W^\mu W_\mu \,s\equiv
  \left(-\theta\frac{\sqrt{2}}{v}\sin(\beta-\alpha)
  -\psi\frac{\sqrt{2}}{v}\cos(\beta-\alpha)\right) m_W^2W^\mu W_\mu\,s.    
\end{equation}
The width of $s\to W^+W^-$ decay from both contributions is
given by the formula  
\begin{multline} \label{eq:sWW}
  \Gamma\left({s}\rightarrow
  WW\right)=\frac{1}{16\pi}\frac{m_W^4}{m_s}
  \left[2C^2_{sWW_T}\left(6-4\frac{m_s^2}{m_W^2}
    +\frac{m_s^4}{m_W^4}\right)-\right.  
    \\\left. -12C_{sWW_T}C_{{s}WW_L}\left(1-\frac{m_s^2}{2m_W^2}  
    \right)+C^2_{{s}WW_L}\left(3-\frac{m_s^2}{m_W^2}
    +\frac{m_s^4}{4m_W^4}\right) \right]\sqrt{1-\frac{4m_W^2}{m_s^2}}\,.  
\end{multline}
Similarly for $Z$-boson one obtains 
\begin{multline}
  \Gamma\left({s}\rightarrow ZZ\right)=\frac{1}{8\pi}\frac{m_Z^4}{m_s}
  \left[2C^2_{sZZ_T}\left(6-4\frac{m_s^2}{m_Z^2}+\frac{m_s^4}{m_Z^4}\right)
    -\right.\\\left.-12C_{sZZ_T}C_{{s}ZZ_L}\left(1-\frac{m_s^2}{2m_Z^2}
    \right)+C^2_{{s}ZZ_L}\left(3-\frac{m_s^2}{m_Z^2}
    +\frac{m_s^4}{4m_Z^4}\right) 
    \right]\sqrt{1-\frac{4m_Z^2}{m_s^2}}. 
\end{multline}
The respective trilinear coefficients for $Z$-bosons are
\begin{equation}
  C_{sZZ_T}=-\frac{M_1\sin^2{\theta_W}+M_2\cos^2{\theta_W}}{2\sqrt{2}F},
\end{equation}
\begin{equation}
  C_{{s}ZZ_L}=-\frac{\theta}{v\sqrt{2}}\sin(\beta-\alpha)
  -\frac{\psi}{v\sqrt{2}}\cos(\beta-\alpha).  
\end{equation}
The decay widths of sgoldstino to fermions $s\rightarrow
f\bar{f}$ are small as compared to the decay widths to bosons. Also 
note that sgoldstino can decay into pair of gravitinos, however
the corresponding partial width is strongly suppressed in the
interesting case of $m_s\ll \sqrt{F}$.   
In summary, the total sgoldstino decay width can be found as the
following sum over all the considered decay channels
\begin{equation}
  \Gamma_{tot}=\Gamma\left(s\rightarrow
  gg\right)+\Gamma\left(s\rightarrow \gamma
  \gamma\right)+\Gamma\left(s\rightarrow \gamma
  Z\right)+\Gamma({s}\rightarrow
  hh)+\Gamma\left({s}\rightarrow
  WW\right)+\Gamma\left({s}\rightarrow ZZ\right). 
\end{equation}
The total width scales as $1/F^2$ and for $m_{soft},\mu\ll \sqrt{F}$
sgoldstino width is typically considerably smaller than its mass.

\begin{figure}[t] 
  \begin{center}
    \includegraphics[width=0.8\textwidth]{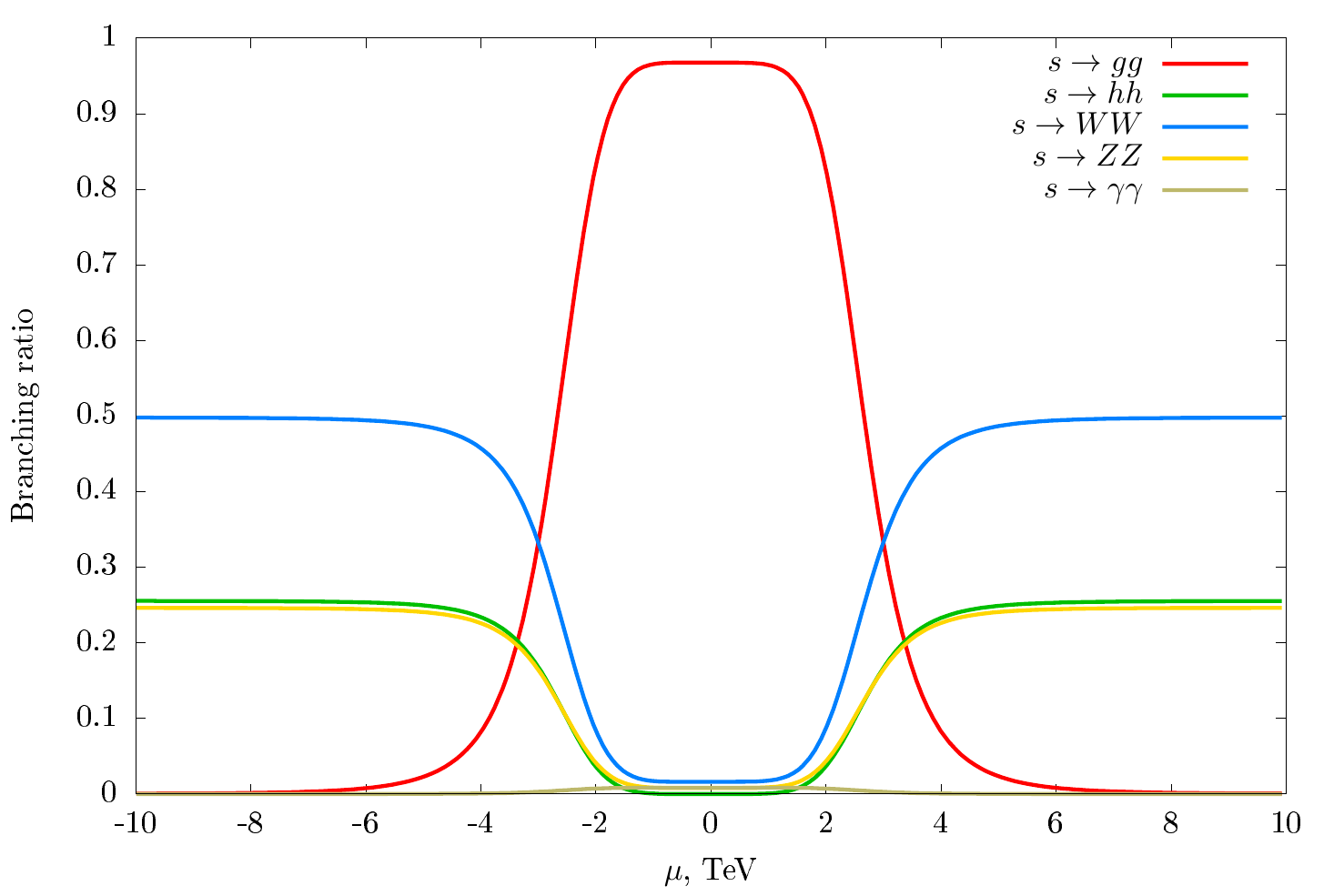}
    \caption{Branching ratios for sgoldstino decays in two gluons
      ($s\rightarrow gg$, red line), two light neutral Higgs bosons
      ($s\rightarrow hh$, green line), two W-bosons ($s\rightarrow
      WW$, blue line), two Z-bosons ($s\rightarrow ZZ$, yellow line),
      two photons ($s\rightarrow \gamma \gamma$, brown line) versus
      parameter $\mu$. Here the model parameters are fixed as follows:
      $\tan \beta=10$, $m_s=1$ TeV, $m_A=5$ TeV, $M_1=M_2=1$ TeV,
      $M_3=3$ TeV, $\sqrt{F}=20$ TeV, K-factor for $gg$ channel equals
      $K_d=1.6$.} 
    \label{fig:br}			
  \end{center}
\end{figure}

Let us discuss how the hierarchy between different sgoldstino decay
modes depends on its mixing with other part of the Higgs sector. It is
known that at large $\sqrt{F}$ even small mixing with the Higgs bosons
can change the patterns of dominant sgoldstino decay
modes~\cite{Astapov:2014mea}. In our analysis we concentrate mostly on
the decoupling regime, 
i.e. when $m_A, m_H\gg m_h$, which corresponds to $\sin{\alpha} \approx
-\cos{\beta}$ and $\cos{\alpha}\approx\sin{\beta}$. 
In this limit, along with $m_{soft}, \mu\ll \sqrt{F}$ which is required for
small values of the mixing angles, the scalar sgoldstino of the mass
at TeV scale mixes mainly with the lightest Higgs boson. The
corresponding mixing angle determines the parameter $X$
in~\eqref{eq:X} which becomes  
\be
\label{eq:X_decoupling}
X=-v\biggl(\frac{g_1^2M_1+g_2^2M_2}{2}v^2 \cos^2{2\beta}+ 2\mu^3\sin{2\beta}
\biggr)\,.
\ee
For the part of the model parameter space relevant for our study the last
term in this expression 
dominates. To illustrate the hierarchy between the sgoldstino decay
modes, we show in Fig.~\ref{fig:br}
their branching ratios as functions of the parameter $\mu$.
Here the values of other model parameters are fixed as $\tan
\beta=10$, $m_A=5$ TeV, $M_1=M_2=1$ TeV, $M_3=3$ 
TeV, $\sqrt{F}=20$ TeV, $K_d=1.6$ and sgoldstino mass is $m_s=1$ TeV. We
observe two different regimes of possible sgoldstino
decays. At relatively small values of $\mu$ and therefore small
sgoldstino-Higgs mixing, sgoldstino decays dominantly into gluons,
whereas the partial widths of the other decay channels and, in 
particular, the decay into two light neutral Higgs bosons, are
suppressed. Note, that most of the previous studies of sgoldstino
phenomenology in high energy collisions assumed gluon dominance in
sgoldstino decays. In the opposite case of large absolute values of
$\mu$, i.e. with a considerable admixture of the Higgs bosons in
sgoldstino, the latter decays mainly into heavy particles of the
electroweak sector, namely, the Higgs $hh$ and massive vector
$W^+W^-$, $ZZ$ bosons. In the decoupling limit the corresponding
partial decay widths are related approximately as 1:2:1
respectively. This regime was discussed in Ref.~\cite{Asano:2017fcp}
in the context of searches for sgoldstino at the LHC. Our present
interest in this study is related to the possibility of 
resonant double Higgs production in this part of the model parameter
space, as the branching ratio of two light neutral Higgs boson 
channel can reach $25\%$. Note that in this regime sgoldstino decays
also into massive vector bosons which can help to test this 
scenario. 

In this calculation we use the approximate formulas for mixing
angles for simplicity. We checked this approximation by computing 
mixing angles using (exact) numerical diagonalization of the mass
matrix. 
The results from approximate formulas agree very well with 
branching ratios obtained numerically.  

\begin{figure}
  \begin{center}
    \begin{subfigure}[t]{0.49\textwidth}
      \begin{center}
	\includegraphics[width=\textwidth]{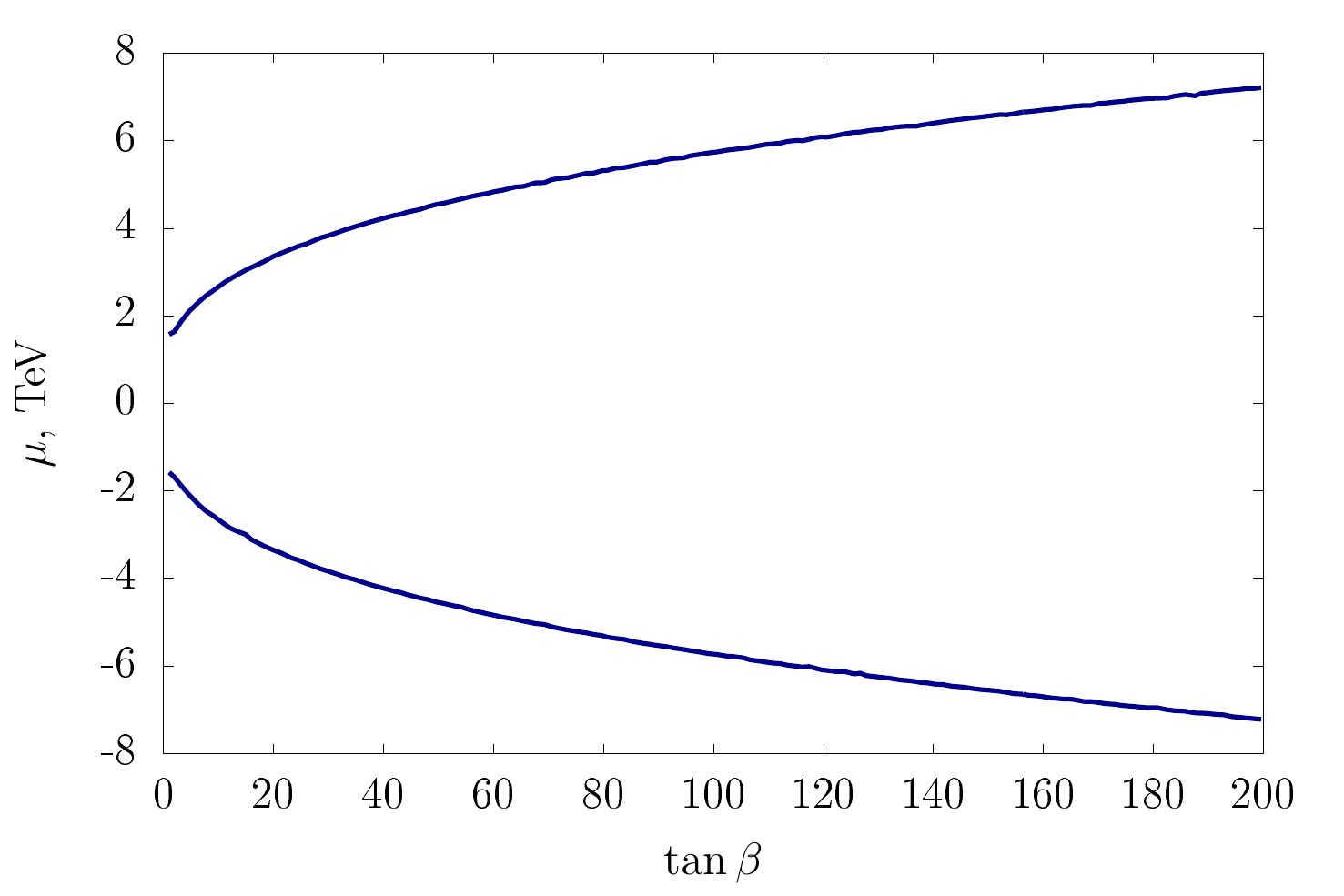}
	\label{fig:mulrtan}
      \end{center}			
    \end{subfigure}
    \begin{subfigure}[t]{0.49\textwidth}
      \begin{center}
	\includegraphics[width=\textwidth]{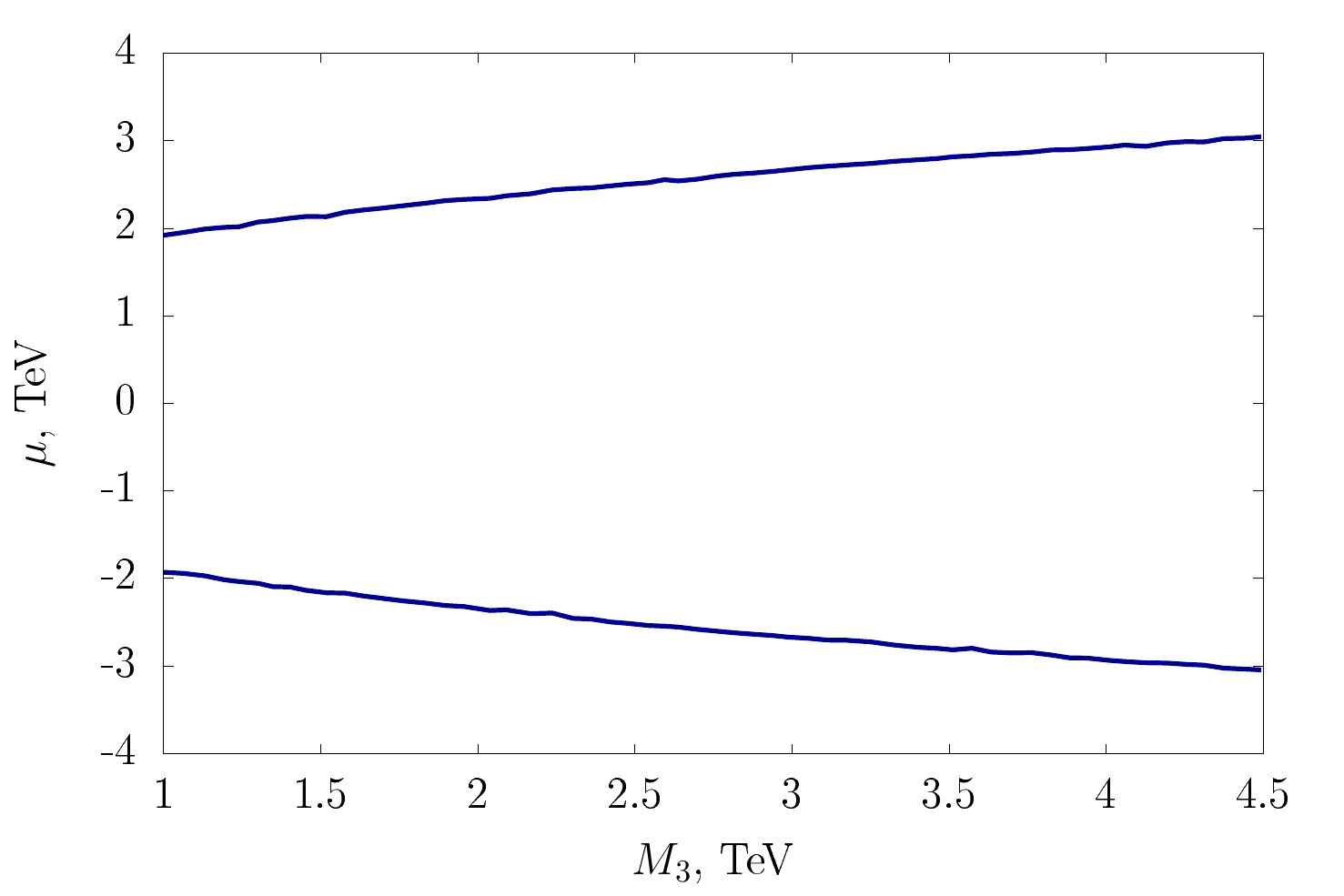}
	\label{fig:mulrM3}
      \end{center}
    \end{subfigure}
    \begin{subfigure}[b]{0.49\textwidth}
      \begin{center}
	\includegraphics[width=\textwidth]{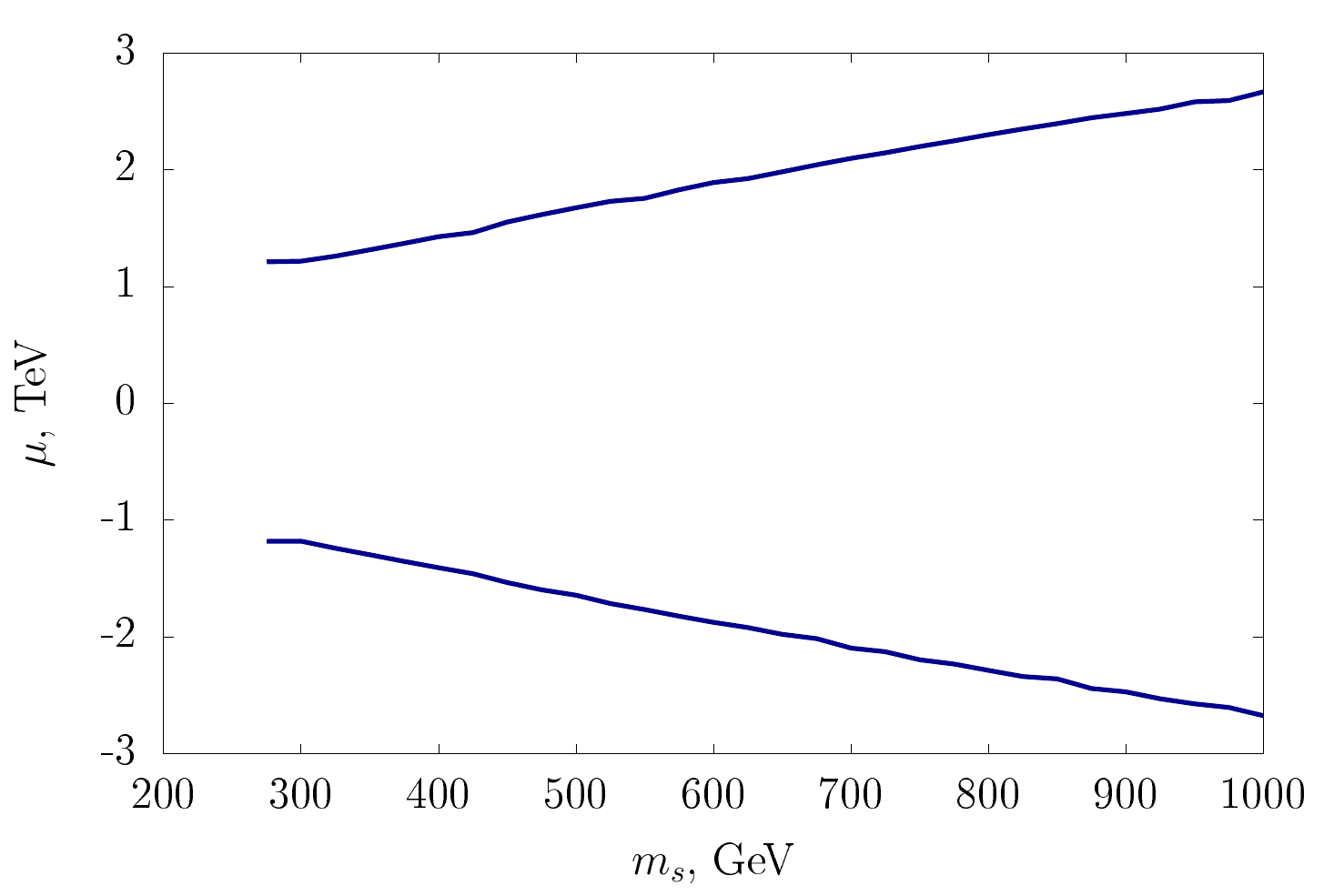}
	\label{fig:mulrms}
      \end{center}
    \end{subfigure}
    \begin{subfigure}[b]{0.49\textwidth}
      \begin{center}
	\includegraphics[width=\textwidth]{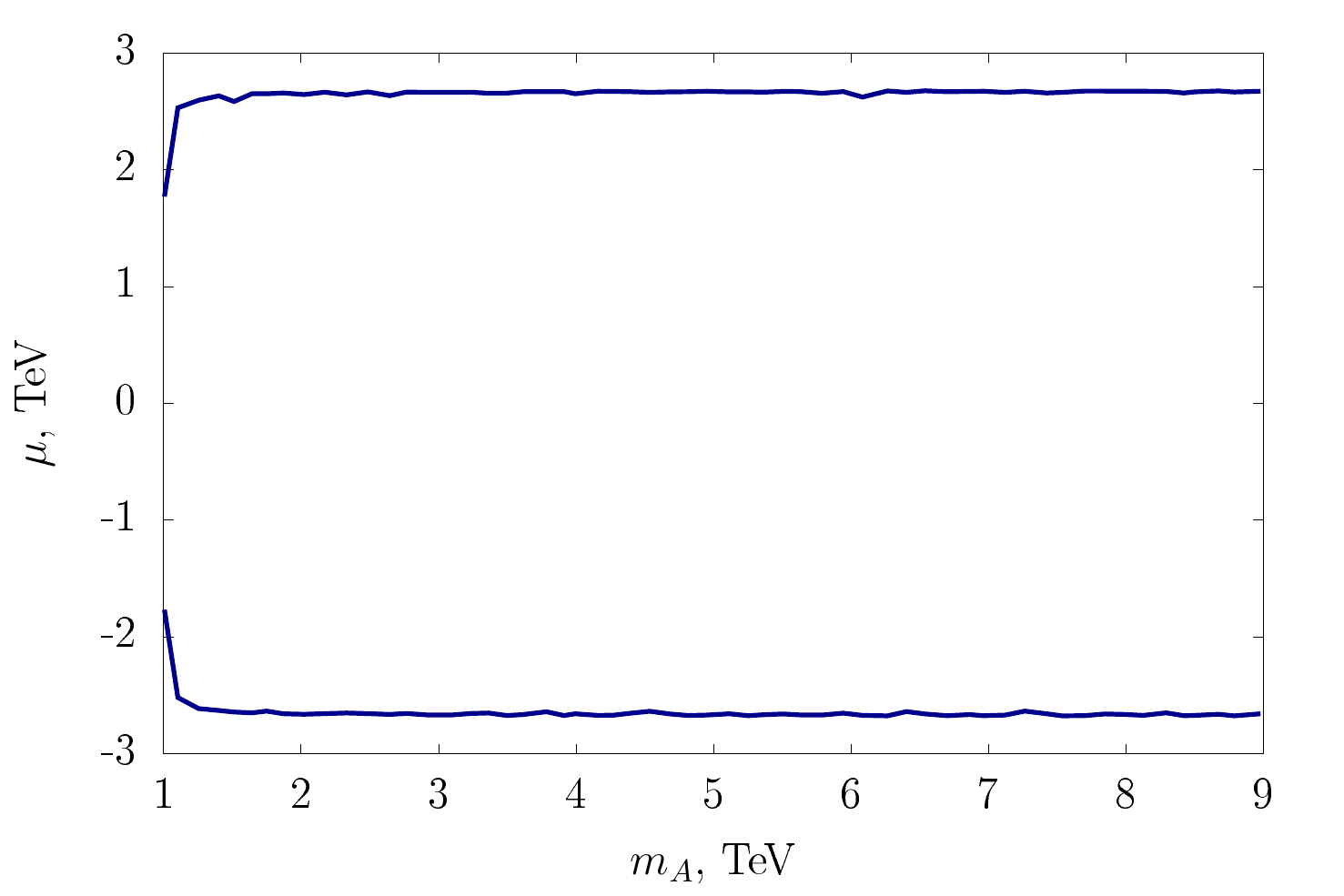}
	\label{fig:mulrma}
      \end{center}
    \end{subfigure}
    \caption{Slices of the conditional boundary $Br(s\rightarrow hh) = 
      0.125$ between gluon dominated and massive boson dominated 
      regimes of sgoldstino decay on the planes $(\tan{\beta},\mu)$
      (upper left panel), $(M_3,\mu)$ (upper right panel), $(m_s,\mu)$ (lower
      left panel) and $(m_A,\mu)$ (lower right panel). Regions between the lines
      on the plots correspond to the gluon dominated sgoldstino decay.
      Other model parameters are fixed as $\tan{\beta}=10$, $m_A=5$~TeV,
      $m_s=1$~TeV, $M_3=3$~TeV, $M_1=M_2=1$~TeV,
      $\sqrt{F}=20$~TeV. K-factor for $gg$ channel is $K_d=1.6$.} 
    \label{fig:mulr}	
  \end{center}
\end{figure}

Let us study the dependence of the transition between gluon
dominated and heavy boson dominated regimes of sgoldstino decays on
the model parameters. As a conditional boundary between these two
regimes we take $Br(s\rightarrow hh)=0.125$ and in Fig.~\ref{fig:mulr}
we show slices of this boundary on the planes $(\tan{\beta},\mu)$,
$(M_3,\mu)$, $(m_s,\mu)$ and $(m_A,\mu)$ assuming other parameters to
be fixed. Regions between the lines on the plots correspond to
smaller $|\mu|$ and therefore to the gluon dominated sgoldstino decay.
Outside these regions sgoldstino decays mostly to $hh$, $W^+W^-$ and $ZZ$ and
the ratio between their partial widths approaches 1:2:1 at larger
$|\mu|$. We see that larger values of $\tan{\beta}, M_3, m_s$ push the
boundary of the region with gluon dominated sgoldstino decay to larger
values of $\mu$. A nontrivial dependence of the boundaries on $m_A$ is
observed only for its small values, when the decoupling regime gets violated.

Let us present analytical estimates for the dependence of the
boundaries between different regimes on the model parameters. At large mixing angle the leading
contribution to the sgoldstino decay width into two W-bosons~\eqref{eq:sWW} for
$m_s\gg m_W$ is
\begin{equation}
  \Gamma(s\rightarrow WW)\sim \frac{1}{4\pi}
  \frac{\theta^2}{8v^2} m_s^3\,.
\end{equation}
The vector bosons dominate in sgoldstino decays if
$\Gamma(s\rightarrow gg) \ll \Gamma(s\rightarrow WW)$ or 
\begin{equation}
  \abs{\frac{2M_3 v}{F\theta}+\frac{\alpha_s A_t}{3\pi}} \ll 1\,,
\end{equation}
where we denote  $A\left({4m_t^2}/{m_s^2}\right) = A_t$ for brevity.
For all values of sgoldstino mass in the range 200--1000 GeV we find 
$\abs{A_t}\leq1.9$ and $\alpha_s A_t/(3\pi)\ll 1$. Therefore,
the condition of dominating decays of sgoldstino to vector bosons is
$\abs{\theta} \gg \theta_{cr}$, where  
\begin{equation} \label{eq:thetacr}
  \theta_{cr}=2\sqrt{2}\frac{M_3v}{F}.
\end{equation}

%%%%%%%%%%%%%%%%%%%%%%%%%%%%%%%%%%%%%%%%%%%%%%%%%%%%%%%%%%%%%%%%%%%%%%%%%%%%
\section{Phenomenology of the resonant sgoldstino signature}
\label{higgsprod}
	
In this Section we discuss sgoldstino production at the LHC yielding 
the resonant signatures in $hh$, $W^+W^-$ and $ZZ$ final states. 
Our primary interest here is in double-Higgs mode, which we start
from; the other two we investigate similarly. Also recall, that
sgoldstino partial decay widths for these double-boson final states are related
as 1:2:1 for $\theta\gg\theta_{cr}$, and so the modes are compatible. 
To calculate the cross section 
of the 
resonant di-Higgs production in the process $pp\rightarrow
s\rightarrow hh$, 
we work in
the narrow width approximation for sgoldstino which means 
that the cross section for production of particles in a final state
$fin$ is found as a product of the sgoldstino production cross
section (at the given energy of protons) and the corresponding
branching ratio ${\rm Br}(s\rightarrow fin)$. 

\begin{figure}
  \begin{center}
    \begin{subfigure}{0.49\textwidth}
      \includegraphics[width=0.99\textwidth]{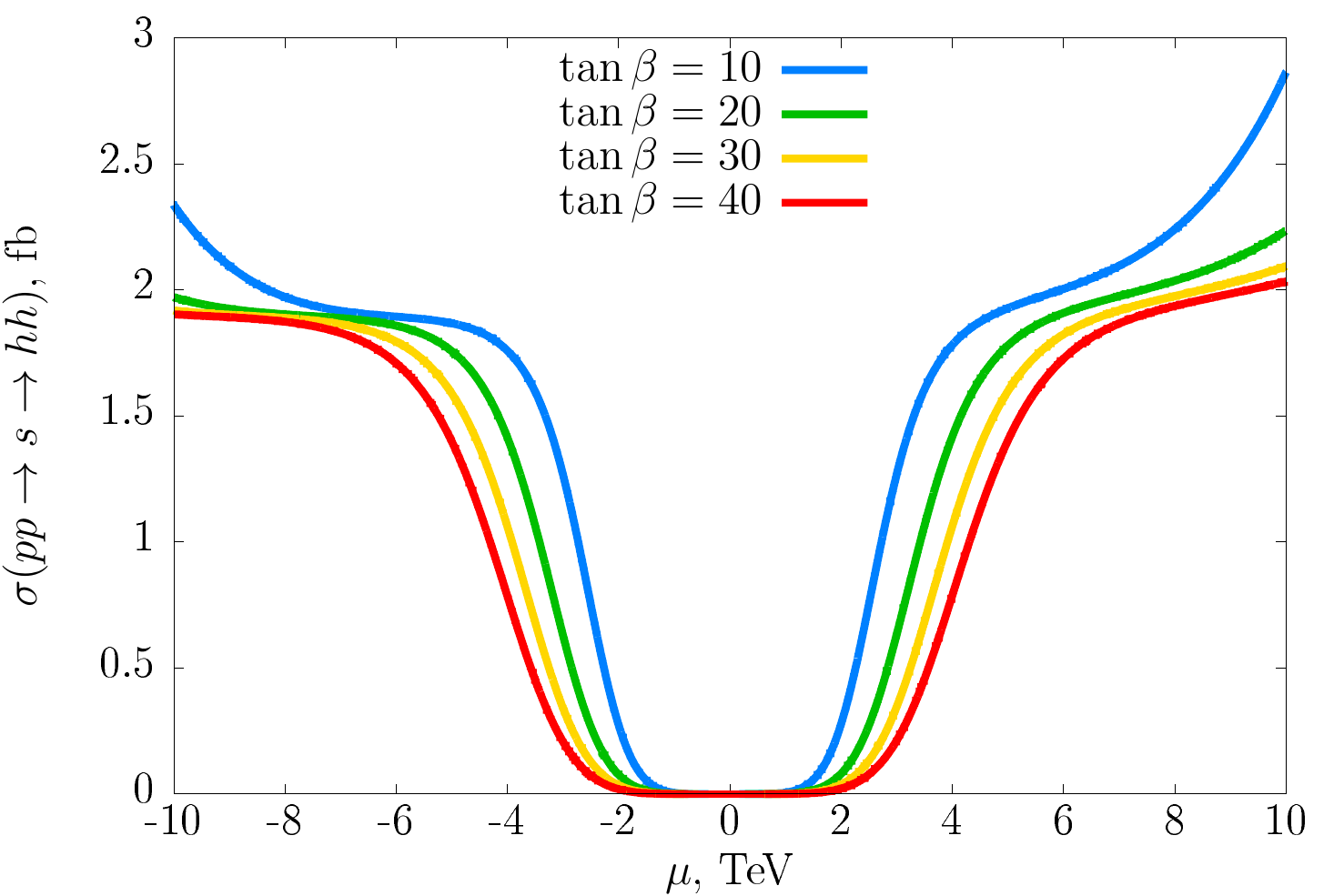}
      \caption{}
      \label{fig:tan}
    \end{subfigure}
    \begin{subfigure}{0.49\textwidth}
      \includegraphics[width=0.99\textwidth]{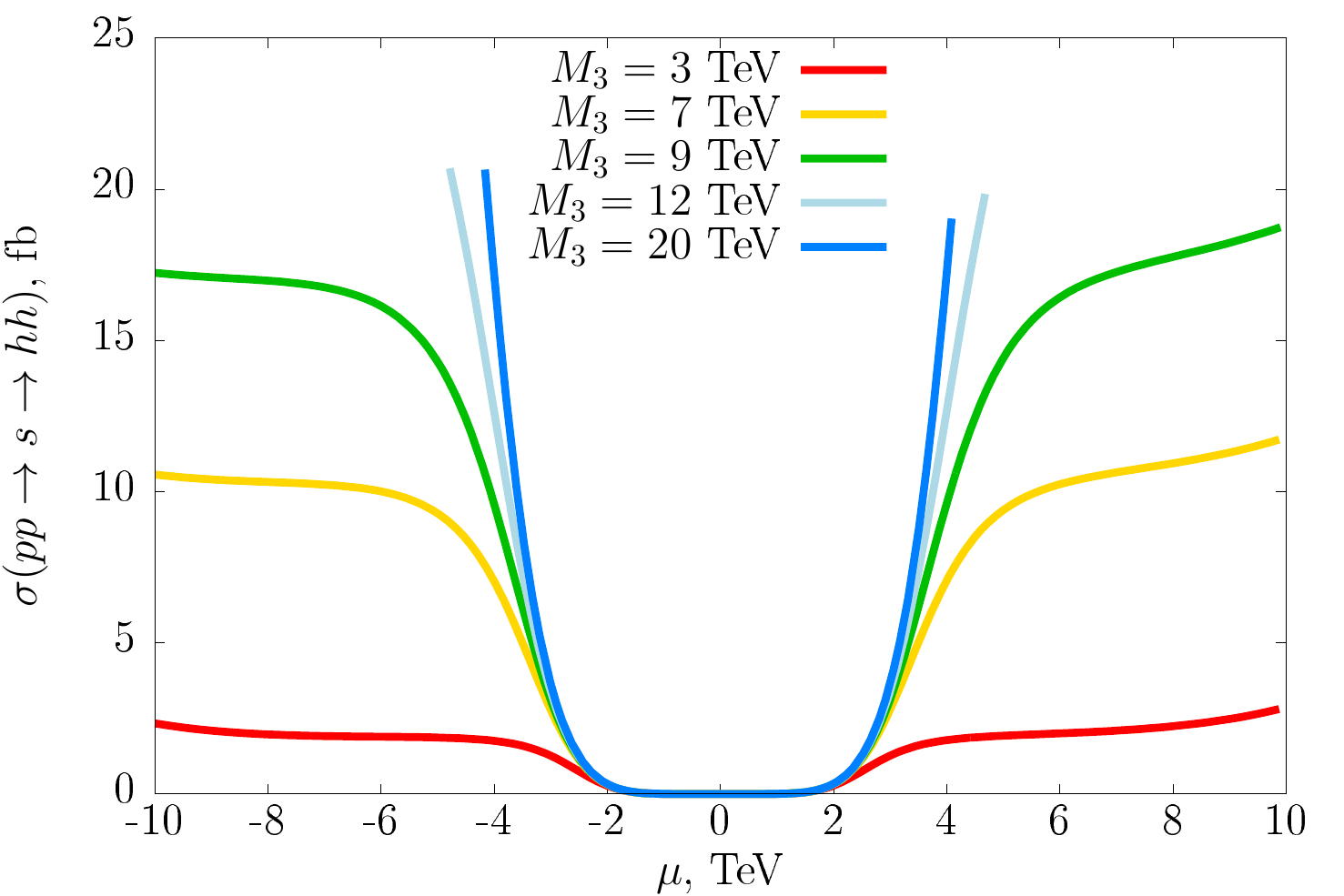}
      \caption{}
      \label{fig:M3}
    \end{subfigure}
    \caption{\label{13TeV} Resonant di-Higgs production cross section of the process
      $pp\rightarrow s\rightarrow hh$ as a function of $\mu$ for
      different values of $\tan{\beta}$ (a) and $M_3$ (b), calculated
      for $\sqrt{S}=13$ TeV.
      Other parameters are fixed as $m_s=1$ TeV, $m_A=5$ TeV, $M_1=M_2=1$
      TeV, $\sqrt{F}=20$ TeV, $K_p=2$, $K_d=1.6$;  $M_3=3$~TeV~(a)
    and $\tan{\beta}=10$~(b).}			 
  \end{center}
\end{figure}

The dependence of the resonant double Higgs boson production cross
section on $\mu$ for chosen values of $\tan{\beta}$
and $M_3$ is shown in Figs.~\ref{fig:tan} and~\ref{fig:M3}.
Other parameters are fixed as follows: $m_s=1$~TeV, $m_A=5$~TeV,
$M_1=M_2=1$~TeV, $\sqrt{F}=20$~TeV, $K_p=2$, $K_d=1.6$. 
We see that for small values of $|\mu|$ the cross section of this process is
relatively small due to a suppression of the decay channel $s\to hh$. 
At the same time, the cross section is increased for larger $|\mu|$,
i.e. for larger mixing between sgoldstino and the lightest Higgs
boson~\eqref{eq:X_decoupling} which corresponds to transition to the
regime of sgoldstino decay to massive vector bosons. It is in 
this regime the resonant cross sections of $pp\to s\to W^+W^-$ and
$pp\to s\to ZZ$ are directly related to that of di-Higgs production. 
As it can be seen from Fig.~\ref{fig:tan}, with the increase of
$\tan{\beta}$, the value of $\mu$, at which the double Higgs
production is enhanced, also increases. The sgoldstino production
cross section grows with $M_3$ and therefore the double Higgs
production cross section also increases.

At large values of $M_3$ the
points with large absolute values of $\mu$ in Fig.~\ref{fig:M3} are
excluded by existing experimental data obtained by ATLAS and CMS
collaborations. The search for a scalar resonance that decays to two
$W$-bosons  was performed
in~\cite{Aaboud:2017fgj,Aaboud:2017gsl}. Scalar 
resonance decaying into $ZZ$ was  studied in~\cite{Aaboud:2017itg}; see
also Refs.\,\cite{Sirunyan:2017hsb} for \cite{Aaboud:2017uhw} the case
of $Z$-boson and photon in the final state and
Ref.\,\cite{Aaboud:2017yyg} for a pair of photons.
We also take into account current upper limits on the production cross
section of a scalar which decays into two Higgs bosons which are 
presented in~\cite{Aaboud:2018knk, Aaboud:2018ftw, Aaboud:2018ewm,
  Aaboud:2018zhh, 
  Sirunyan:2018two, Aaboud:2018ksn, Sirunyan:2019quj}.
We use the whole set of the above experimental constraints to find 
the phenomenologically viable region in the model parameter
space. In Fig.~6 we show the results for phenomenologically viable models
only. 
	
One can notice that the cross section in Fig.~\ref{fig:tan} is nearly
constant at intermediate absolute values of $\mu$ but starts growing
again at its larger absolute values. This behavior is due 
to the change in the dominant contribution to the sgoldstino effective
coupling to gluons. Namely, according to~\eqref{eq:X}
and~\eqref{eq:theta} the larger 
values of $\abs{\mu}$ correspond to larger $\abs{X}$ and
$\abs{\theta}$. In this regime, the cross section of sgoldstino
production~\eqref{eq:sgldprod} is determined mostly by the sgoldstino
mixing with the Higgs boson. Parametrically this occurs when
\begin{equation} \label{eq:ineq}
  \frac{M_3}{F}\ll \frac{\alpha_s \abs{\theta} \abs{A_t}}{6\pi v},
\end{equation}
or $\abs{\theta}\gg \theta'_{cr}$ with $\theta'_{cr}\equiv 6\pi v M_3/(\alpha_s
|A_t| F)$. Thus for $\theta$ larger than the critical value
$\theta'_{cr}$, the sgoldstino production cross section increases with
the growth of $\abs{\mu}$. For $\theta_{cr}\ll\theta'_{cr}$,
this cross section does not depend on the value of the mixing angle
and it is determined fully by values of sgoldstino mass, $M_3$ and $F$.   
	
\begin{figure}
    \begin{center}
    \includegraphics[width=0.9\textwidth]{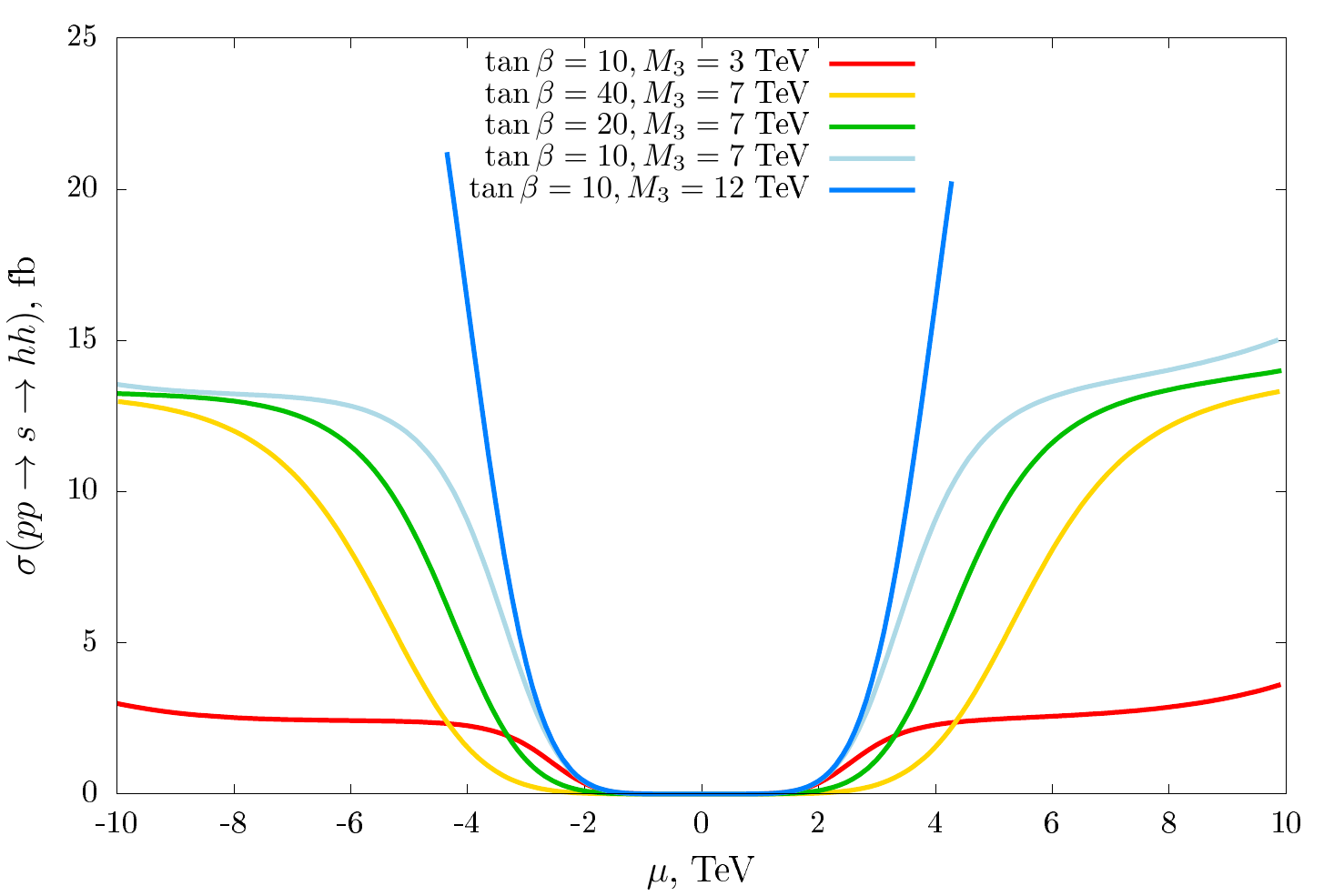}
    \caption{Resonant di-Higgs production cross section of the process
    $pp\rightarrow s\rightarrow hh$ calculated for $\sqrt{S}=14$ TeV.
    The model parameters are chosen as $m_s=1$ TeV, $m_A=5$ TeV, $M_1=M_2=1$ TeV, $\sqrt{F}=20$
    TeV, $K_p=2$, $K_d=1.6$.}
    \label{fig:predicthh}
  \end{center}
\end{figure}

\begin{figure}
	\begin{center}
		\begin{subfigure}{0.9\textwidth}
			\begin{center}
				\includegraphics[width=0.99\textwidth]{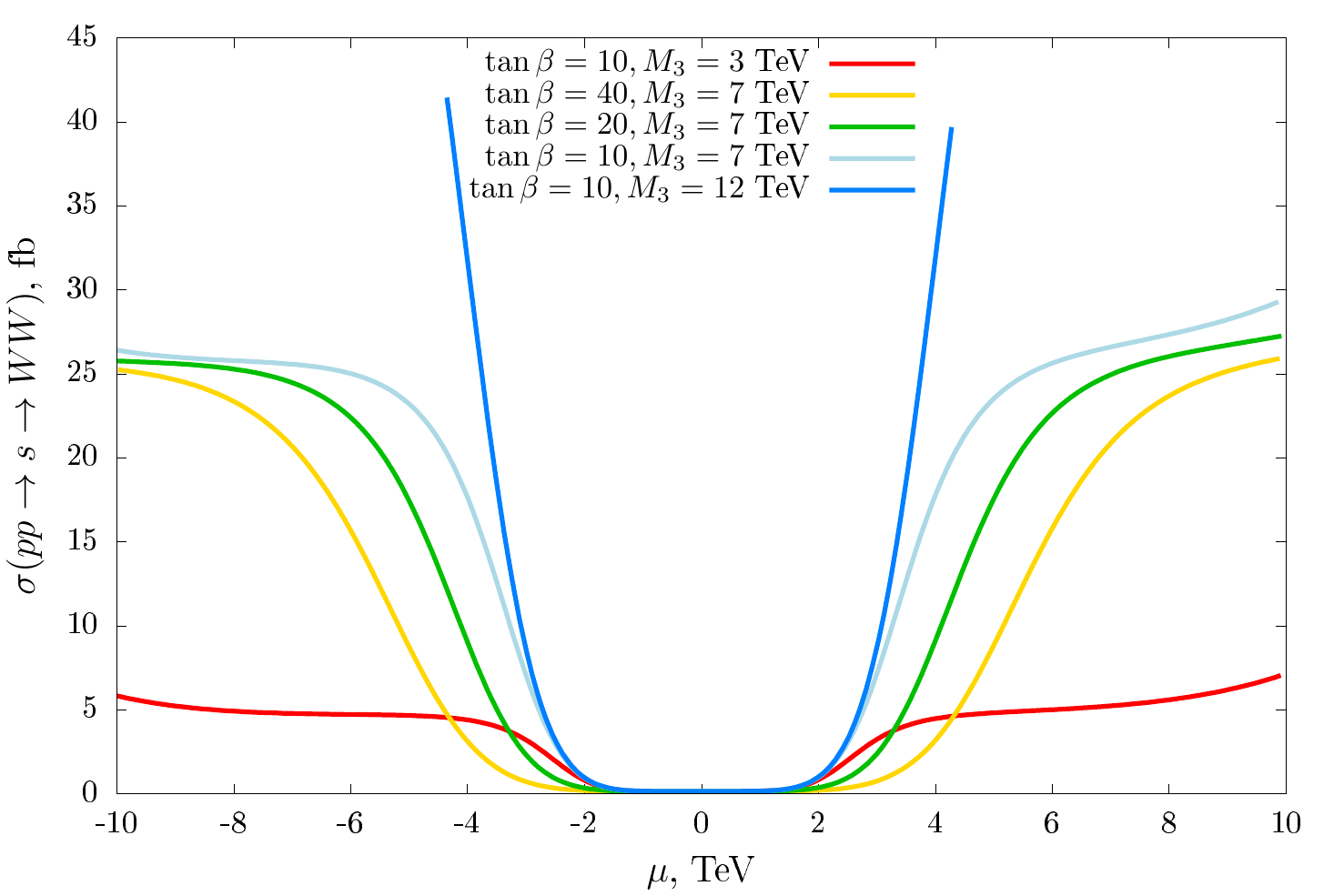}
			\end{center}
		\end{subfigure}
		\begin{subfigure}{0.9\textwidth}
			\begin{center}
				\includegraphics[width=0.99\textwidth]{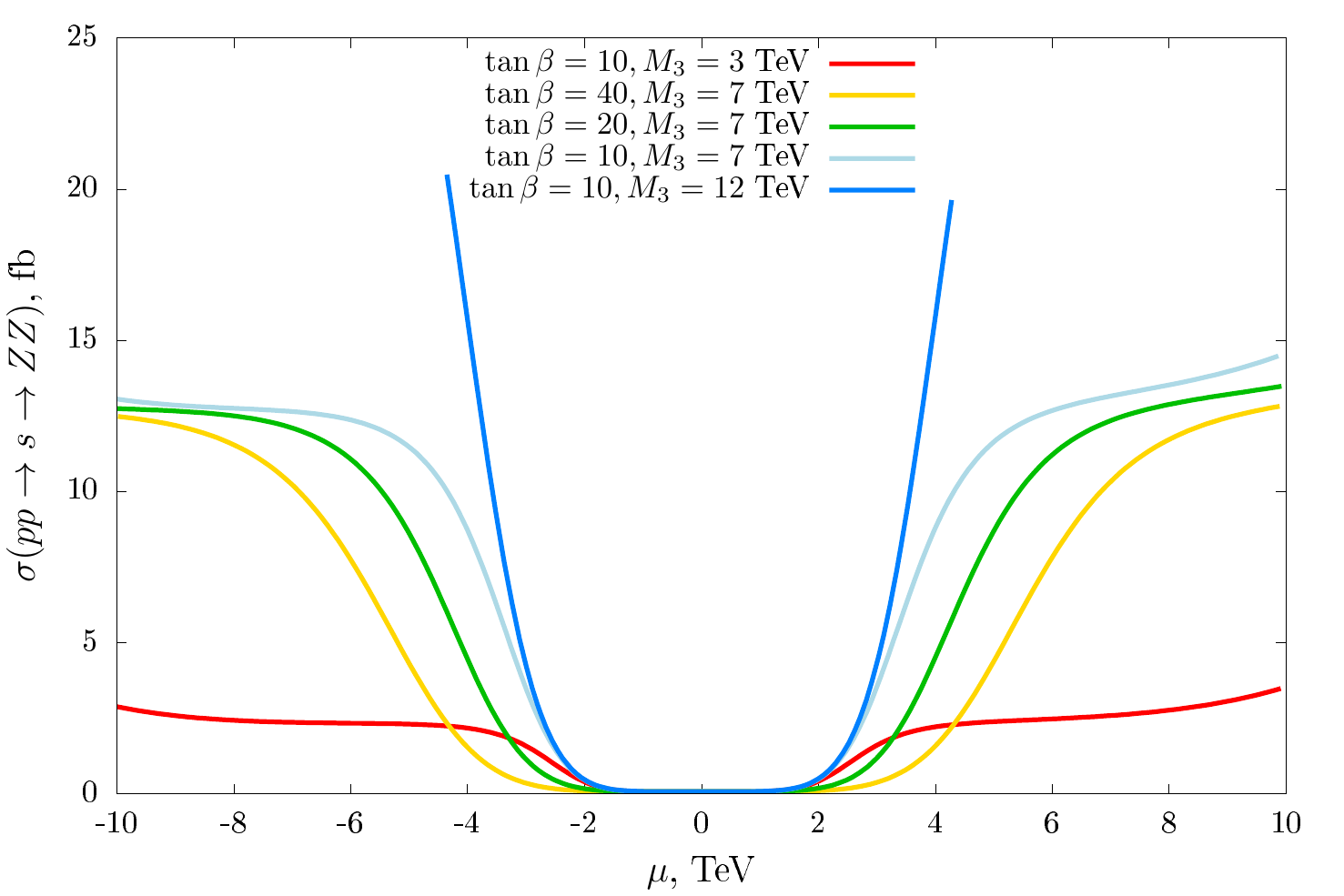}
			\end{center}
		\end{subfigure}
		
		\caption{Resonant production cross section of the processes
			$pp\rightarrow s\rightarrow WW$ (upper panel), $pp\rightarrow s\rightarrow ZZ$ (lower panel) calculated for $\sqrt{S}=14$ TeV.
			The model parameters are chosen as $m_s=1$ TeV, $m_A=5$ TeV, $M_1=M_2=1$ TeV, $\sqrt{F}=20$ TeV, $K_p=2$, $K_d=1.6$.}
		\label{fig:predictWWZZ}
	\end{center}
\end{figure}

In Fig. \ref{fig:predicthh}
we show the similar dependencies of the cross section $pp\to s\to hh$
as in Fig.~\ref{13TeV} but calculated for $\sqrt{S}=14$~TeV.
The cross section is depicted for five different
combinations of $\tan\beta$ and $M_3$ and for the 
same set of other parameters. Similarly, in Fig. \ref{fig:predictWWZZ} the cross sections of the processes $pp\to s\to WW$ and $pp\to s\to ZZ$ are calculated for $\sqrt{S}=14$~TeV. Let us note again that at
$\theta>\theta_{cr}$ the following relations between the resonant sgoldstino
production cross sections 
\be
\sigma(pp\to s\to W^+W^-)\approx 2\sigma(pp\to s\to ZZ)\approx
2\sigma(pp\to s \to hh)
\ee
are valid.

\begin{figure}
  \begin{center}
    \includegraphics[width=0.9\textwidth]{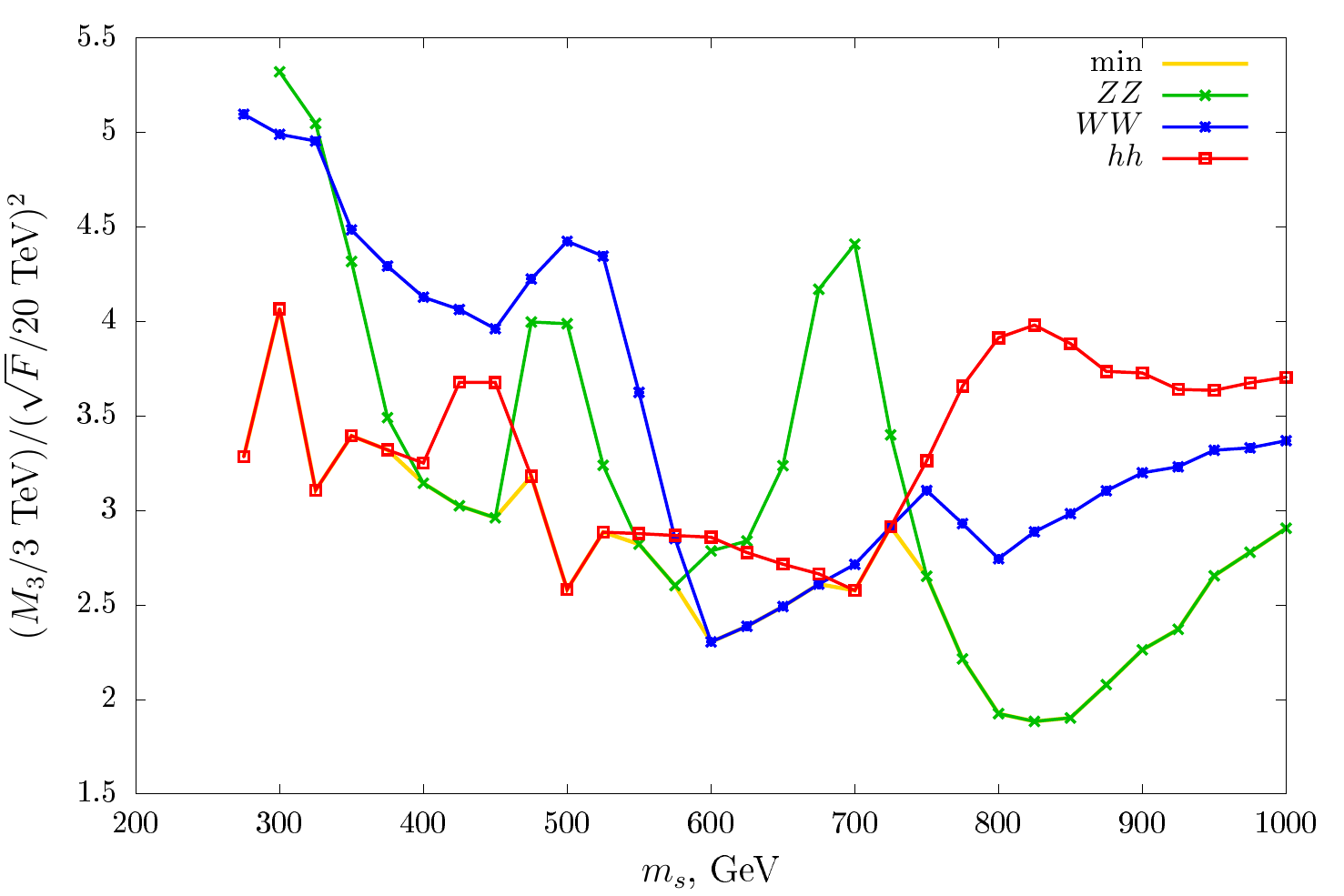}
    \caption{Upper limits at 95$\%$~CL on ratio $M_3/F$ obtained from the experimental
      data \cite{Aaboud:2017fgj, Aaboud:2017gsl, Aaboud:2017itg},
      \cite{Aaboud:2018knk, Aaboud:2018ftw, Aaboud:2018ewm,
        Aaboud:2018zhh, Sirunyan:2018two, Aaboud:2018ksn,
        Sirunyan:2019quj} for $\sqrt{S}=13$\,TeV.
      Constraints obtained from the searches for a heavy scalar 
      resonance decaying into a pair of $Z$-bosons, $W$-bosons and
      Higgs bosons are shown in green, blue and red, respectively. The
      yellow line connects points of joint constraint on $M_3/F$
      (minimum upper limit).}  
    \label{fig:uplim}
  \end{center}
\end{figure}

In the rest of this Section we assume the regime of sgoldstino
decaying dominantly into $hh$, $W^+W^-$ and $ZZ$ with the relation
1:2:1 between their partial widths. In this regime, the
results of searches for neutral scalar resonances by the ATLAS and the
CMS collaborations at $\sqrt{S}=13$~TeV discussed above can be used to
constrain sgoldstino production cross section and corresponding
parameters of the model. Assuming $\theta\gg \theta_{cr}$ (which
corresponds to the regime 1:2:1 for sgoldstino decays) as well as
$\theta\ll\theta'_{cr}$ (meaning that the first term
in the expression~\eqref{eq:sigma0} for the sgoldstino production cross
section dominates) one obtains that the cross section depends on the
combination $M_3/F$ only. Hence the experimental data can be used to
constrain this quantity.
\begin{equation}
  \left[\frac{M_3/3\text{ TeV}}{(\sqrt{F}/20\text{ TeV})^2}
    \right]^{max} =
  \sqrt{\frac{\sigma_{prod}^{max}}{\sigma'_{prod}}}\leq\sqrt{\frac{\sigma^{max}_{XX}}{\sigma'_{prod}
      Br(s\rightarrow XX)}}, 
\end{equation}
here $\sigma'_{prod}$ is the sgoldstino production cross section for
$M_3=3$ TeV, $\sqrt{F}=20$ TeV, and $X$ runs through $\left\lbrace h, W, Z
\right\rbrace $.  
The obtained constraints are presented in Fig.~\ref{fig:uplim}.
In the case of not very heavy sgoldstinos ($m_s\lsim 600$ GeV) the most
rigorous constraints are given by the decay mode $s \rightarrow
hh$.   

\begin{figure}
  \begin{center}
    \includegraphics[width=0.9\textwidth]{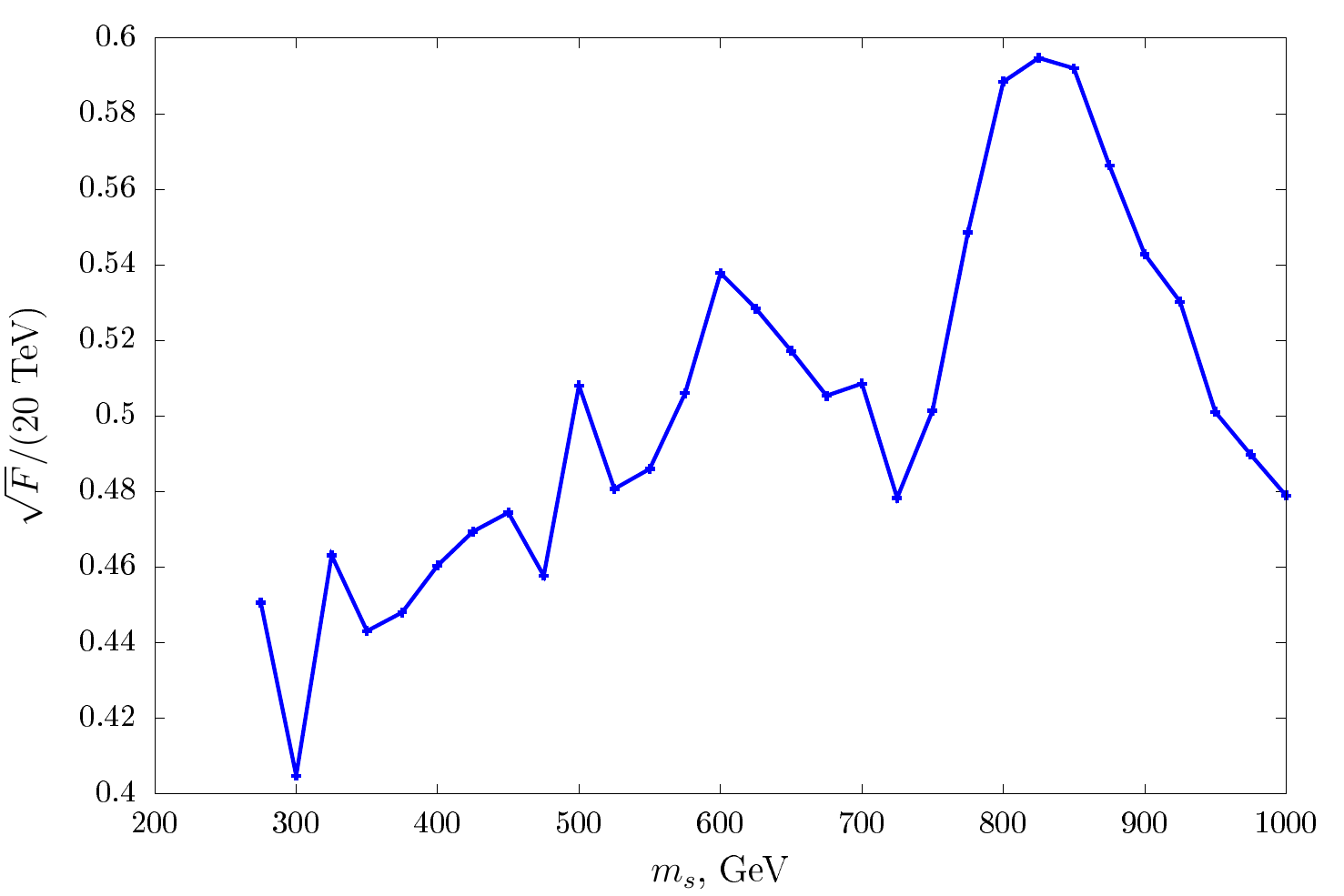}
    \caption{Lower limit on $\sqrt{F}$ obtained using the upper limit
      on $M_3/F$ (Fig.\,\ref{fig:uplim}) and a conservative lower
      bound $M_3\gsim 2$~TeV from experimental data \cite{Kim:2019vcp,
        Sirunyan:2019mbp, Sirunyan:2019hzr, Sirunyan:2018ell,
        Sirunyan:2018psa, Aaboud:2018zeb, Aaboud:2018doq} at
      $\sqrt{S}=13$ TeV.} 
    \label{fig:lowlim}
  \end{center}
\end{figure}

Using lower limits on $M_3$ from the searches for gluino in models with
gauge-mediated supersymmetry breaking along with  the upper bound on
$M_3/F$ presented above one can obtain  constraints  on the supersymmetry
breaking scale $\sqrt{F}$. In most of the experimental
studies~\cite{Kim:2019vcp, Sirunyan:2019mbp, Sirunyan:2019hzr,
  Sirunyan:2018ell, Sirunyan:2018psa, Aaboud:2018zeb, Aaboud:2018doq} 
lower limits on gluino mass were obtained for several simplified
models and the presented experimental bound varies in the range
1.8--2~TeV. Adopting for the estimate the conservative lower bound
$(M_3)_{LL}>2$~TeV and using the constraints on $M_3/F$ from $\sqrt{S}=13$
TeV data shown in Fig.~\ref{fig:uplim}, we apply
\begin{equation}
  F(m_s)\geq\frac{(M_3)_{LL}}{(M_3/F)_{UL}(m_s)},
\end{equation}
to find lower bounds on $\sqrt{F}$ depending on the mass of the scalar
sgoldstino, which are shown in Fig.~\ref{fig:lowlim}.
We find that in the considered part of the model parameter space the
bounds on $\sqrt{F}$ varies in $8-12$~TeV range.  
	
\section{Conclusions}
\label{conclude}

To summarize, in this work we have investigated the impact of
sgoldstino-Higgs mixing on sgoldstino decays and production in 
proton-proton collisions. We have observed two different regimes in
sgoldstino decays. At the small mixing sgoldstino decays dominantly into
gluons while at the large one it decays into the lightest Higgs bosons and
vector bosons, $W^+W^-$ and $ZZ$, with the relation between the
corresponding branching ratios as 1:2:1. In this region of the model
parameter space there can be enhanced resonant di-Higgs
production. Using current experimental data from the LHC experiments 
we obtained constraint on the ratio $M_3/F$. With the current
conservative constraint $M_3\gsim 2$~TeV we placed the bounds on the
supersymmetry breaking scale $\sqrt{F}$ varying in $8-12$~TeV range for
sgoldstino of masses 200--1000~GeV and in the regime of massive boson
dominance in its decays. Since $\sqrt{F}$ can be treated as a scale of
SUSY breaking, the estimate of its lower limit may be useful for a
theoretical consideration of similar low-scale supersymmetry breaking
models with sgoldstino. A characteristic signature of the considered
scenario is the appearance of sgoldstino resonance in di-Higgs,
$W^+W^-$ and $ZZ$ spectra with the relation between corresponding
cross section close to 1:2:1. Searches for the resonant production of
massive boson resonances at HL-LHC are expected to either reveal the
sgoldstino signature or put further
constraints on the models with sgoldstino using updated results of
experimental searches for this process. 

\section*{Acknowledgements}
The work of EK was supported by the grant of ``BASIS'' Foundation no. 19-2-6-16-1.   

\appendix
\section{Trilinear coefficients for a set of sgoldstino-Higgs bosons vertices} \label{trilincoef}
\begin{multline}
  shH\times C_{shH}\equiv
  \frac{shH}{F\sqrt{2}}\left(\frac{v^2}{2}\left(g_1^2M_1+g_2^2M_2\right)
  \left(2\sin{2\alpha}\cos{2\beta}+\cos{2\alpha}\sin{2\beta}\right)+\right.
  \\ \left. +\mu(m_A^2-2\mu^2)\cos{2\alpha}\right),   
\end{multline}
\begin{equation}
  sAA\times C_{sAA} \equiv
  \frac{sAA}{F\sqrt{2}}\left(\frac{v^2}{4}(g_1^2M_1+g_2^2M_2)
  \cos^2{2\beta}-\mu\sin{2\beta}(m_A^2-\mu^2)\right),  
\end{equation}
\begin{equation}
  sH^+H^-\times C_{sH^+H^-}\equiv \frac{sH^+H^-}{F\sqrt{2}}
  \left(-g_2^2M_2v^2(1+\sin{2\beta})-2\mu\left(m_A^2(1+\sin{2\beta})-2\mu^2\right)\right), 
\end{equation}
\begin{equation}
  pAH\times C_{pAH}\equiv \frac{pAH}{F\sqrt{2}}\mu(m_A^2-2\mu^2)\cos{(\alpha-\beta)},
\end{equation}
\begin{equation}
  pAh\times C_{pAh}\equiv \frac{pAh}{F\sqrt{2}}\mu(m_A^2-2\mu^2)\sin{(\beta-\alpha)}.
\end{equation}
	
\section{Trilinear coefficients in the mass basis for a set of 
  sgoldstino-Higgs bosons vertices} \label{trilincoefnew} 
	\begin{equation} \label{eq:shH}
	\frac{\tilde{s}\tilde{h}\tilde{H}}{F}\left(C_{shH}-2C_{hhH}\theta-2C_{hHH}\psi\right), 
	\end{equation}
	\begin{equation} \label{eq:sAA}
	\frac{\tilde{s}\tilde{A}\tilde{A}}{F}\left(C_{sAA}-C_{hAA}\theta-C_{HAA}\psi\right), 
	\end{equation}
	\begin{equation} \label{eq:+-}
	\frac{\tilde{s}\tilde{H}^{+}\tilde{H}^{-}}{F}\left(C_{sH^{+}H^{-}}-C_{hH^{+}H^{-}}\theta-C_{HH^{+}H^{-}}\psi\right), 
	\end{equation}
	\begin{equation} \label{eq:pAH}
	\frac{\tilde{p}\tilde{A}\tilde{H}}{F}\left(C_{pAH}-2C_{HAA}\xi\right), 
	\end{equation}
	\begin{equation} \label{eq:pAh}
	\frac{\tilde{p}\tilde{A}\tilde{h}}{F}\left(C_{pAh}-2C_{hAA}\xi\right), 
	\end{equation}
	where definitions of trilinear coefficients for vertices with one sgoldstino and two Higgs bosons can be found in Appendix \ref{trilincoef}, $C_{hhH}$ and $C_{hHH}$ are introduced in \eqref{eq:hhH}, \eqref{eq:hHH}, the remaining MSSM trilinear coefficients are listed below,
	\begin{equation}
	C_{hH^{+}H^{-}} \equiv \frac{1}{\sqrt{2}}g_2^2v\left(\cos{\alpha}-\sin{\alpha}\right)\left(\cos{\beta}+\sin{\beta} \right),  
	\end{equation}
	\begin{equation}
	C_{HH^{+}H^{-}} \equiv \frac{1}{\sqrt{2}}g_2^2v\left(\cos{\alpha}+\sin{\alpha}\right)\left(\cos{\beta}+\sin{\beta} \right),
	\end{equation}
	\begin{equation}
	C_{hAA} \equiv \frac{1}{2\sqrt{2}}\frac{m_Z^2}{v}\cos{2\beta}\sin{(\alpha+\beta)},
	\end{equation}
	\begin{equation}
	C_{HAA} \equiv
	-\frac{1}{2\sqrt{2}}\frac{m_Z^2}{v}\cos{2\beta}\cos{(\alpha+\beta)}. 
	\end{equation}

\end{document}